\documentclass[numberedappendix]{emulateapj}
\usepackage{natbib}
\usepackage{graphicx}

\def\wid{0.4}
\renewcommand{\baselinestretch}{1.18}

\newcommand\bp{\begin{figure}}
\newcommand\ep{\end{figure}}
\newcommand\bpm{\begin{figure*}}
\newcommand\epm{\end{figure*}}
\newcommand\reffig[1]{Figure \ref{fig:#1}}

\newcommand\Refsec[1]{Section \ref{sec:#1}}
\newcommand\reftbl[1]{Table \ref{tbl:#1}}

\newcommand{\be}{\begin{equation}}
\newcommand{\ee}{\end{equation}}
\newcommand{\bea}{\begin{eqnarray}}
\newcommand{\eea}{\end{eqnarray}}

\newcommand{\degree}{^\circ}

\newcommand{\Fermi}{\emph{Fermi}}

\renewcommand\refname{References and Notes}

\begin{document}

\title{Double gamma-ray lines from unassociated Fermi-LAT sources}

\author{Meng Su\altaffilmark{1,*}, Douglas P. Finkbeiner\altaffilmark{1,2}}

\altaffiltext{1}{ 
  Institute for Theory and Computation,
  Harvard-Smithsonian Center for Astrophysics, 
  60 Garden Street, MS-51, Cambridge, MA 02138 USA } 

\altaffiltext{2}{Center for the Fundamental Laws of Nature,
  Physics Department, 
  Harvard University, 
  Cambridge, MA 02138 USA}

\altaffiltext{*}{mengsu@cfa.harvard.edu}








\begin{abstract}
  Gamma ray emission from dark matter subhalos in the Milky
  Way has long been sought as a sign of dark matter particle
  annihilation.  So far, searches for gamma-ray continuum
  from subhalos have been unsuccessful, and line searches
  are difficult without prior knowledge of the line
  energies.  Guided by recent claims of line emission at 111
  and 129 GeV in the Galactic center, we examine the coadded
  gamma-ray spectrum of unassociated point sources in the
  Second \Fermi-LAT catalog (2FGL) using 3.9 years of LAT
  data. Using the \texttt{SOURCE} event class, we find
  evidence for lines at 111 GeV and 129 GeV with a local
  significance of $3.3\sigma$ based on a conservative
  estimate of the background at $E>135$ GeV.  Other 2FGL
  sources analyzed in the same way do not show line emission
  at 111 and 129 GeV.  The line-emitting sources are mostly
  within 30 degrees of the Galactic plane, although this
  anisotropy may be a selection effect.  If the double-line
  emission from these objects is confirmed with future data,
  it will provide compelling support for the hypothesis that
  the Galactic center line signal is indeed from dark matter
  annihilation.
\end{abstract}


\section{Introduction}
\label{sec:introduction}

Although a wide variety of cosmological and astrophysical
observations provide compelling evidence that nonbaryonic dark
matter constitutes $\sim 80\%$ of the total matter in
the Universe, we still know little about its
nature~\citep[e.g.][]{2012arXiv1205.4882B,Hooper:2007Review,Bertone:2005}.

In many models of weakly interacting massive particle (WIMP) dark matter,
particles
annihilate and/or decay to gamma rays directly or indirectly. 
Gamma-ray photons at energies $E\gg1$ GeV travel in straight lines
without significant energy losses in the local Universe, allowing their
spatial distribution to serve as a tracer of the dark matter distribution.
Regions of high dark matter density such as the Galactic center, galaxy
clusters, and dwarf galaxies have been suggested as possible sources.  In
addition, many DM subhalos in the MW may shine in gamma rays and have no
counterpart at other wavelengths, making them promising sources \citep[see
recently e.g.][and references therein]{2011arXiv1111.2613B,
  2012ApJ...747..121A,SIBYL}

The primary challenge in such searches is to understand the
background from conventional astrophysics well enough to
distinguish a dark matter signal.  A ``smoking-gun''
signal of annihilating dark matter would be the discovery of
one or more gamma-ray lines.  The line(s) could be produced
by dark matter decays or annihilations into two photons, or
two-body final states involving one photon plus a Higgs
boson, Z boson, or other chargeless non-SM particle.  No
plausible astrophysical background can produce a line,
although a narrow feature is possible
\citep[see][]{2012arXiv1207.0458A}.  In most models, the
line flux is suppressed by a loop factor relative to the
continuum, implying it is 2-3 orders of magnitude
fainter~\citep[e.g.][]{Bergstrom:1997}.  Although this is
not true in all
models~\citep[e.g.][]{Bergstrom:1998,Bergstrom:2000,Bertone:2009,Jackson:2010,Cline:2012}, this theoretical prejudice led previous
studies to focus on continuum searches.  However, tentative
evidence for gamma-ray line emission at $\sim$130 GeV toward
the inner Galaxy has been found with 3.3$\sigma$
significance after trials factor\footnote{Also known as the ``look elsewhere effect''}
correction~\citep{weniger:2012}.

\begin{figure*}[ht]
  \begin{center}
    \includegraphics[width=\wid\textwidth]{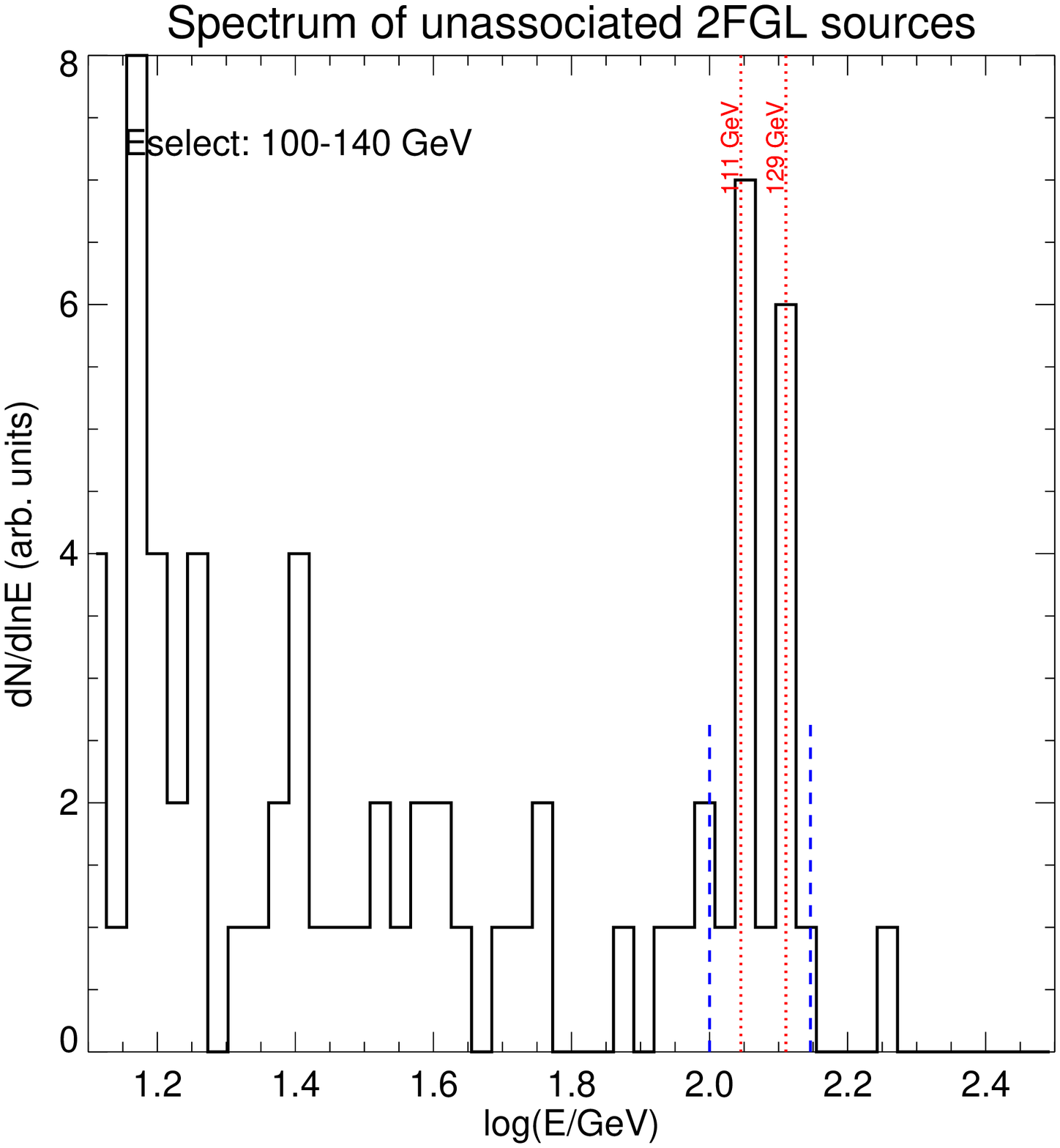}
    \includegraphics[width=\wid\textwidth]{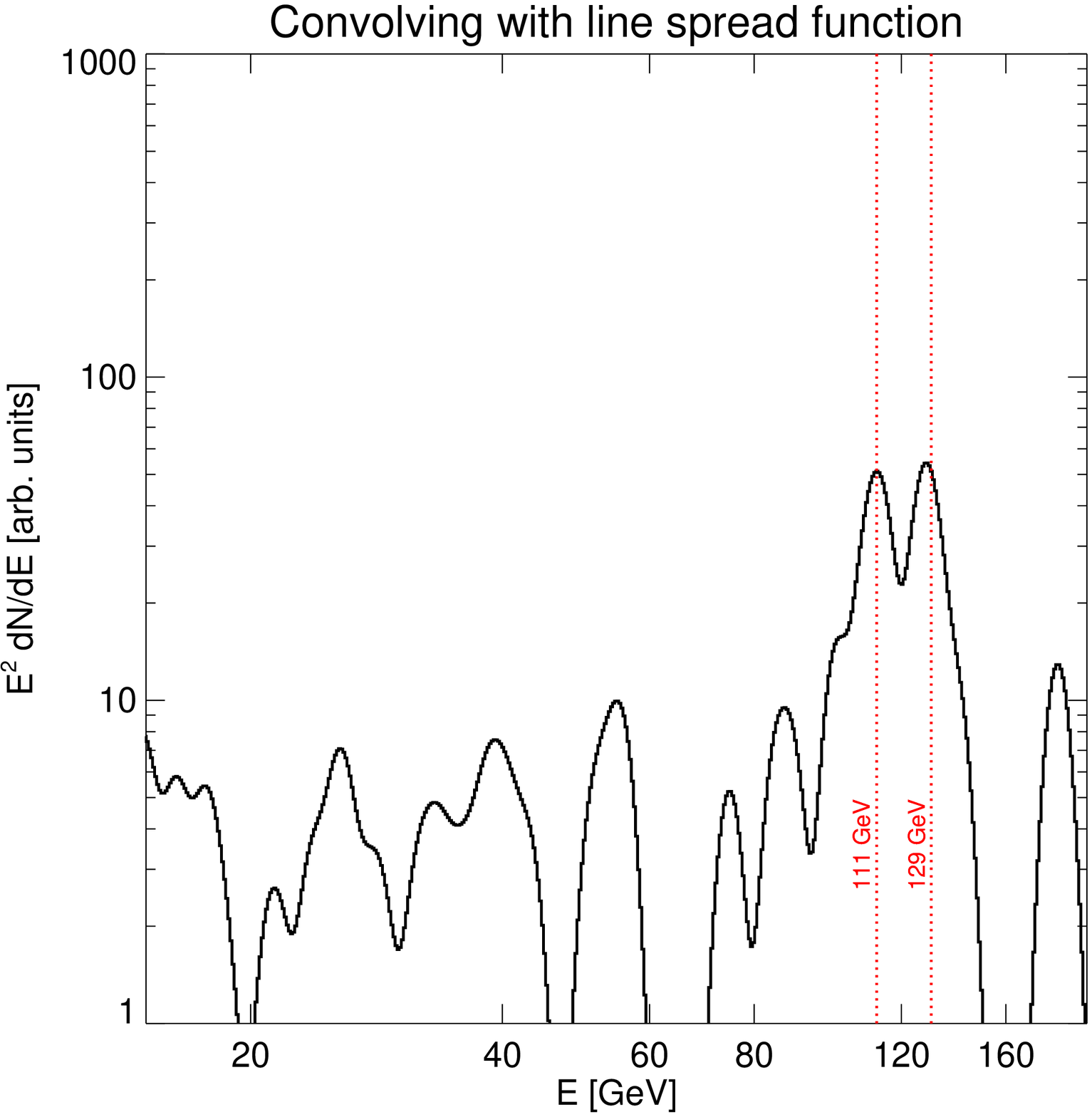}
    \includegraphics[width=\wid\textwidth]{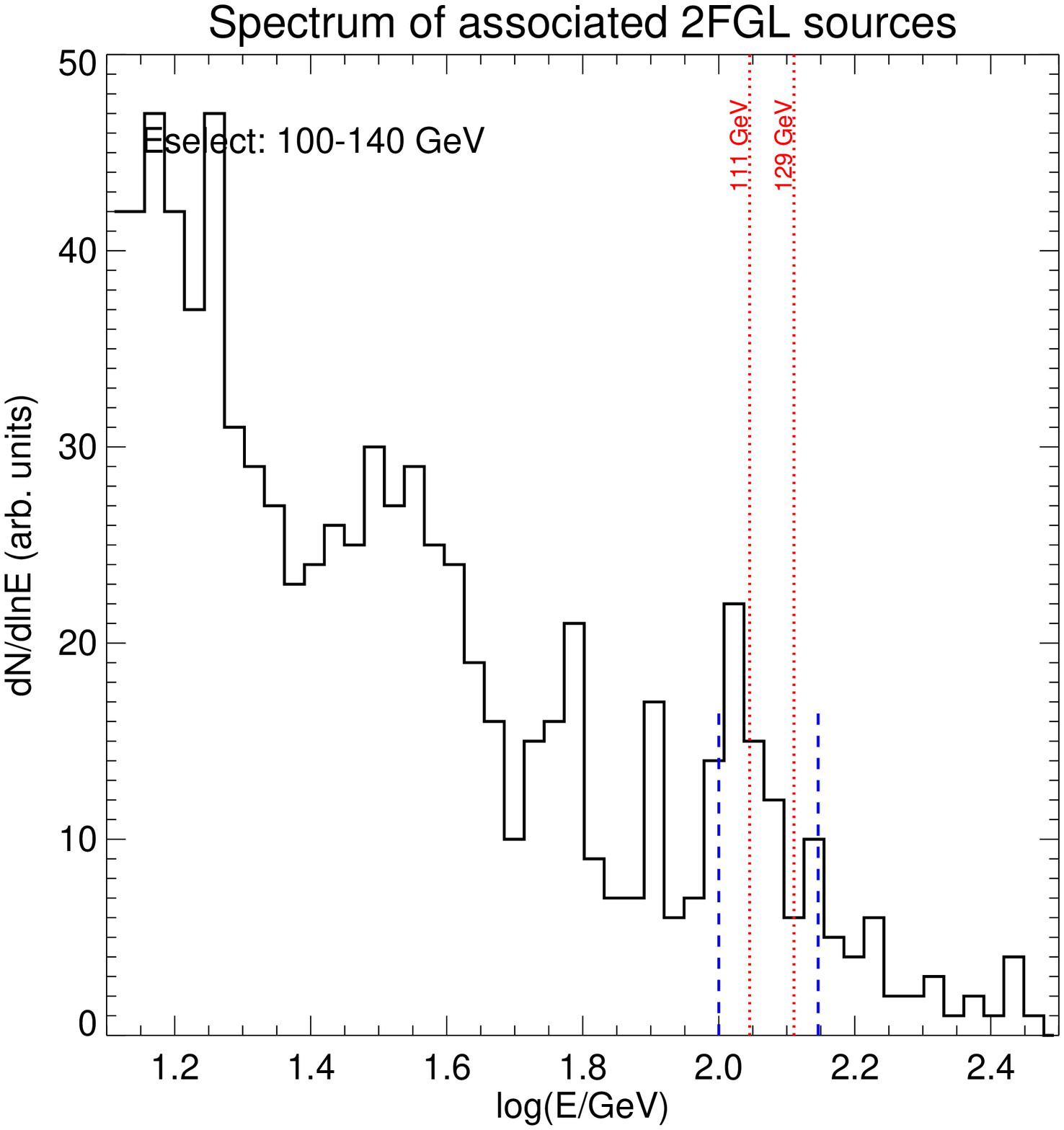}
    \includegraphics[width=\wid\textwidth]{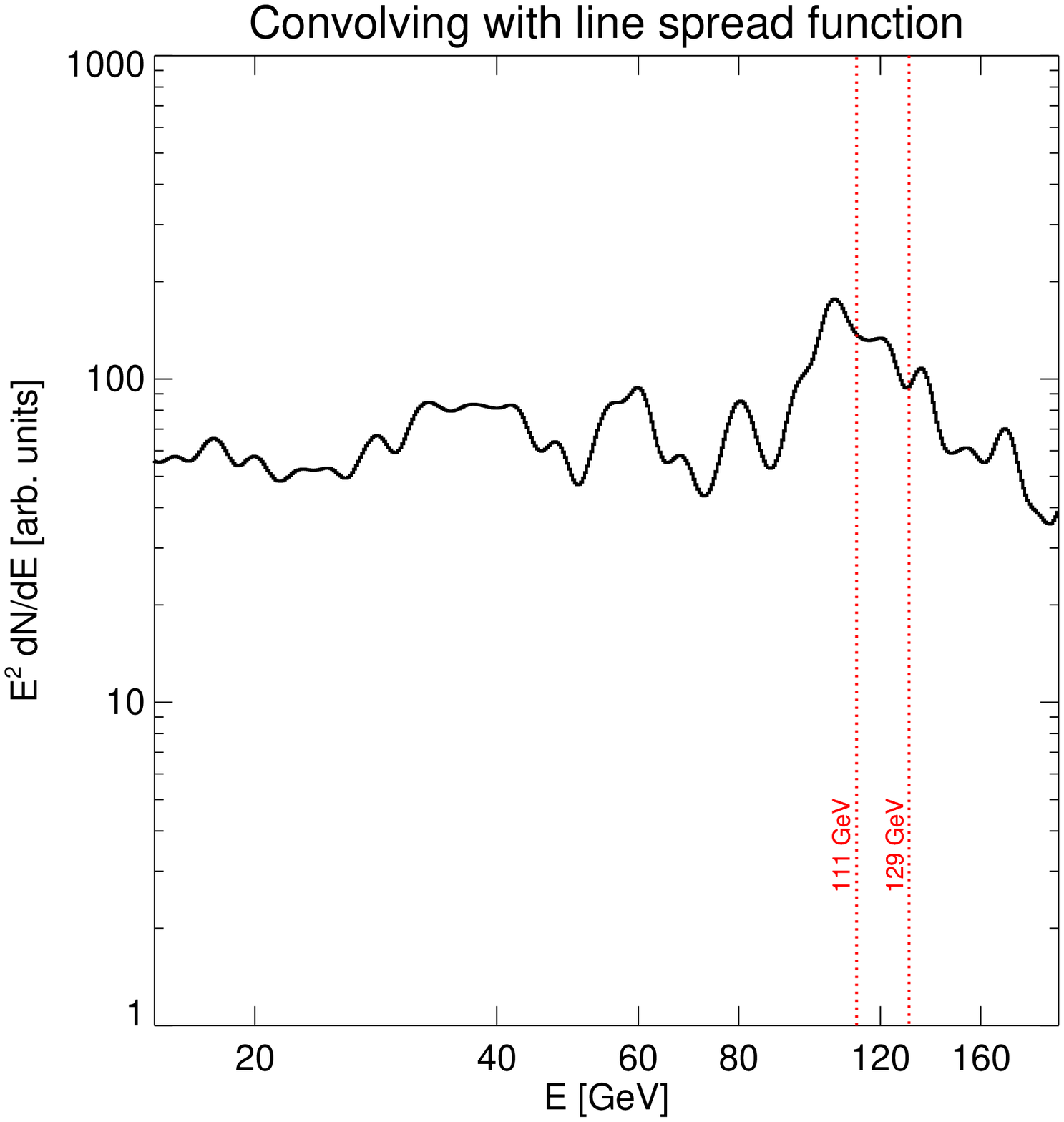}
  \end{center}
  \caption{Upper left: the spectrum of 16 unassociated point sources in the
    2FGL catalog, chosen to have a photon at $E=100-140$ GeV (vertical blue
    lines).  The photons are drawn from the \texttt{SOURCE} event class in 3.9
    years of data, with a match radius of 0.15/0.3$\degree$ for
    \texttt{FRONT/BACK} converting events.  The expected energies (red
    vertical lines) and histogram binning (black) are identical to our
    previous paper on the Galactic center~\citep{linepaper}.  The upper right
    panel is the same spectrum after convolving with the LAT line spread
    function \citep{Rajaraman:2012} and multiplying by $E$.  The lower two
    panels are the same as the upper two panels, but for associated point
    sources.  The associated source spectrum shows a bump between the blue
    lines due to the strong selection bias, but shows no evidence for a
    doublet at the expected energies.}
  \label{fig:fig1}
\end{figure*}

\begin{figure*}[ht]
  \begin{center}
    \includegraphics[width=\wid\textwidth]{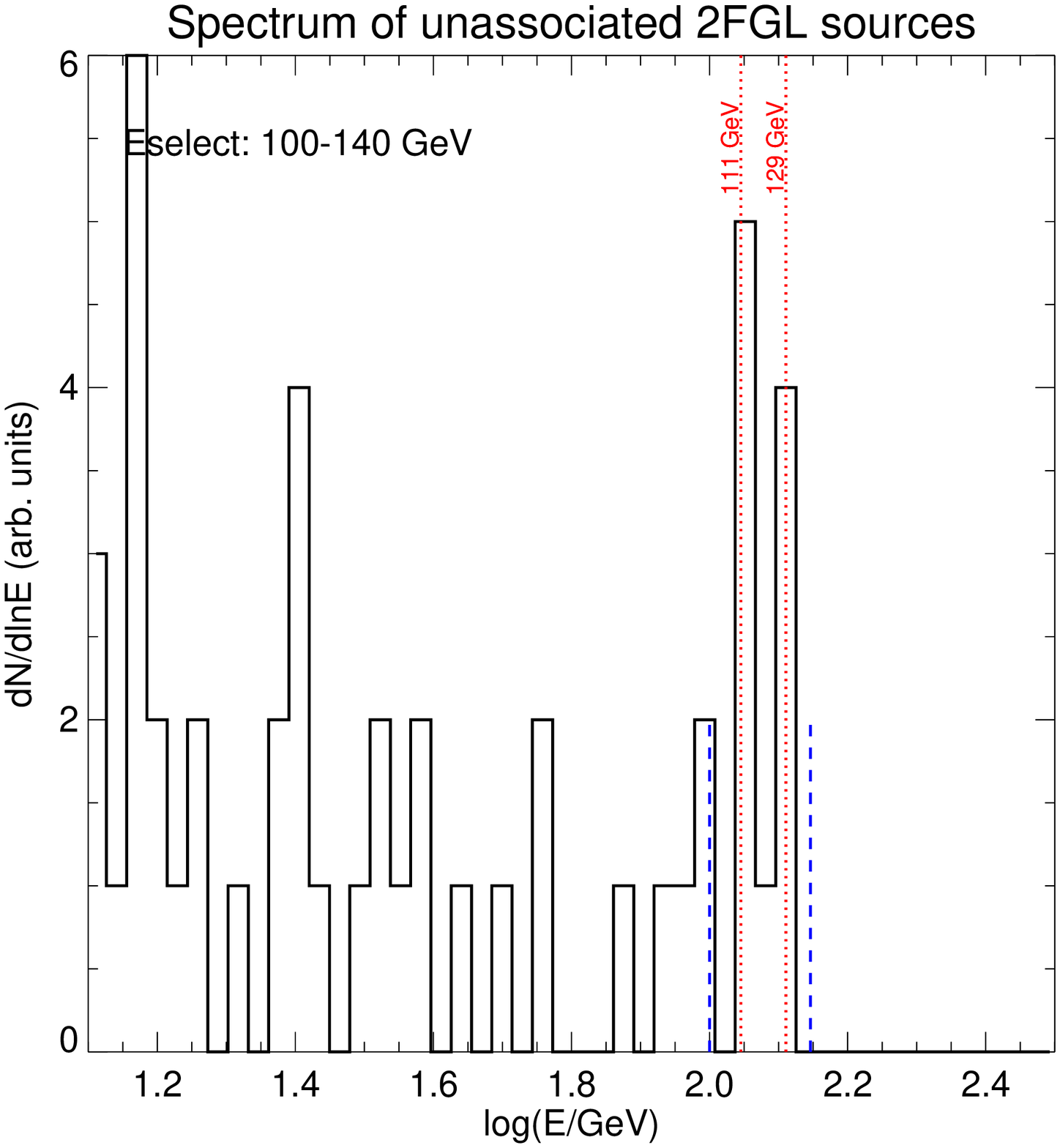}
    \includegraphics[width=\wid\textwidth]{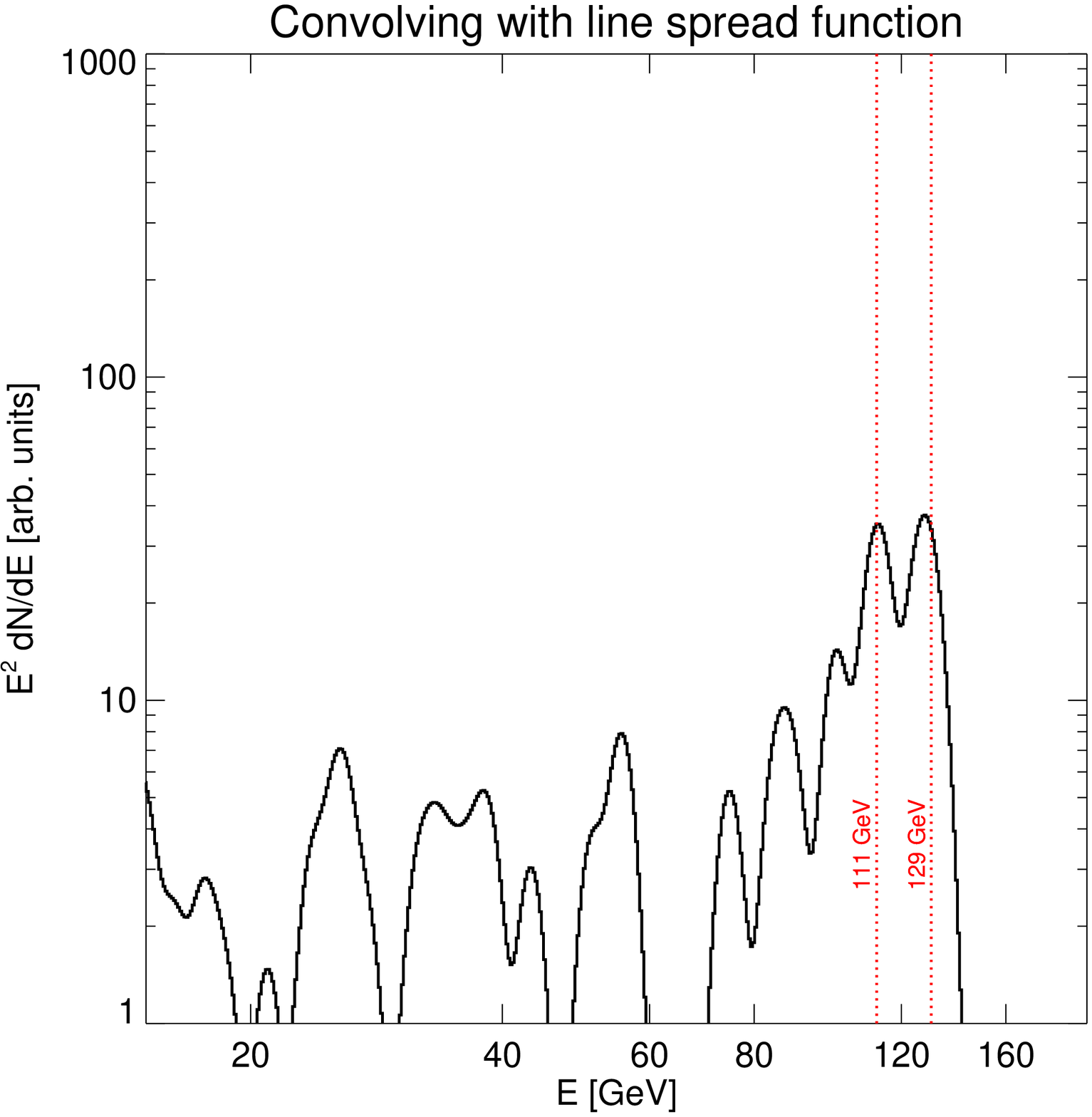}
    \includegraphics[width=\wid\textwidth]{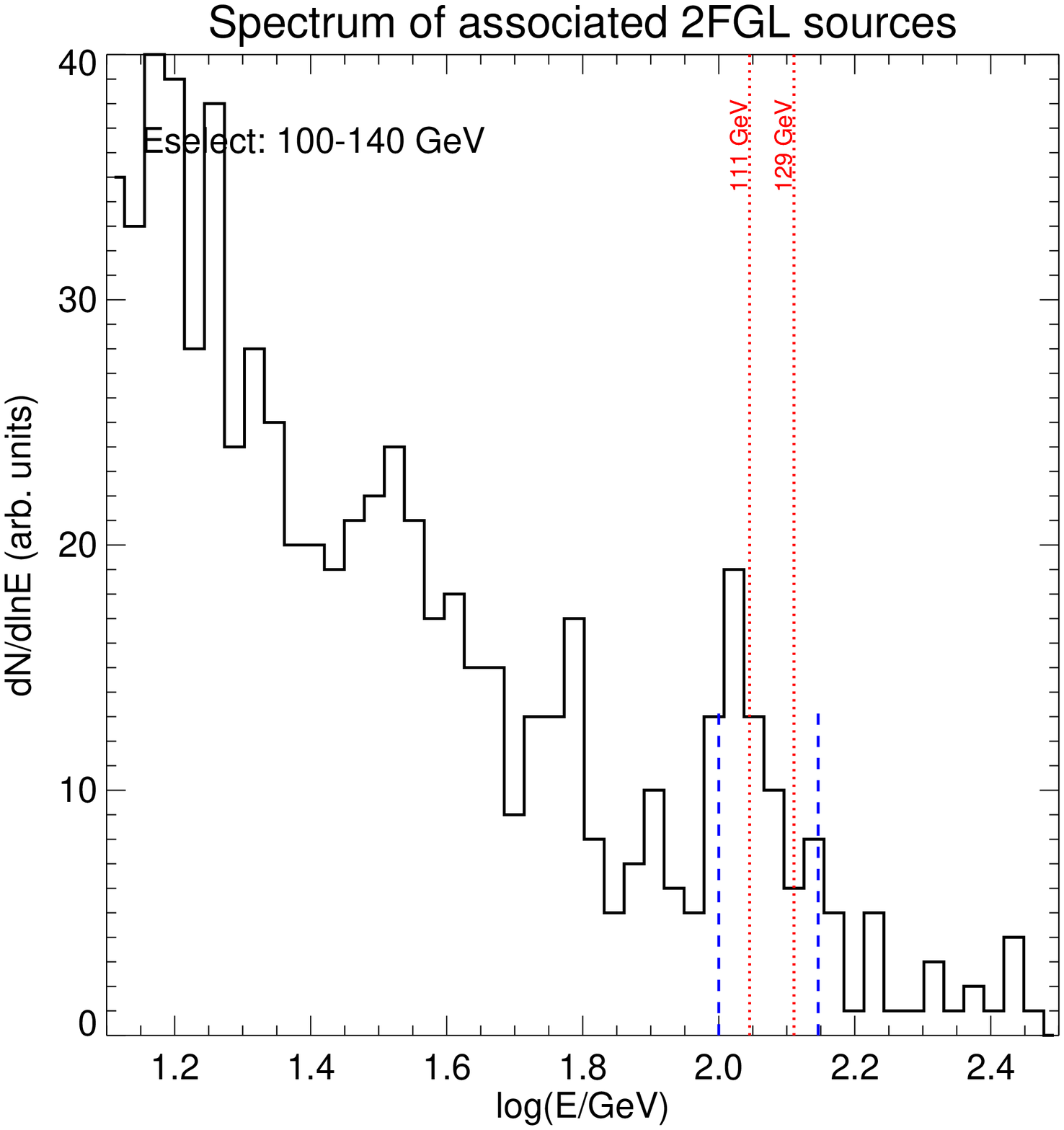}
    \includegraphics[width=\wid\textwidth]{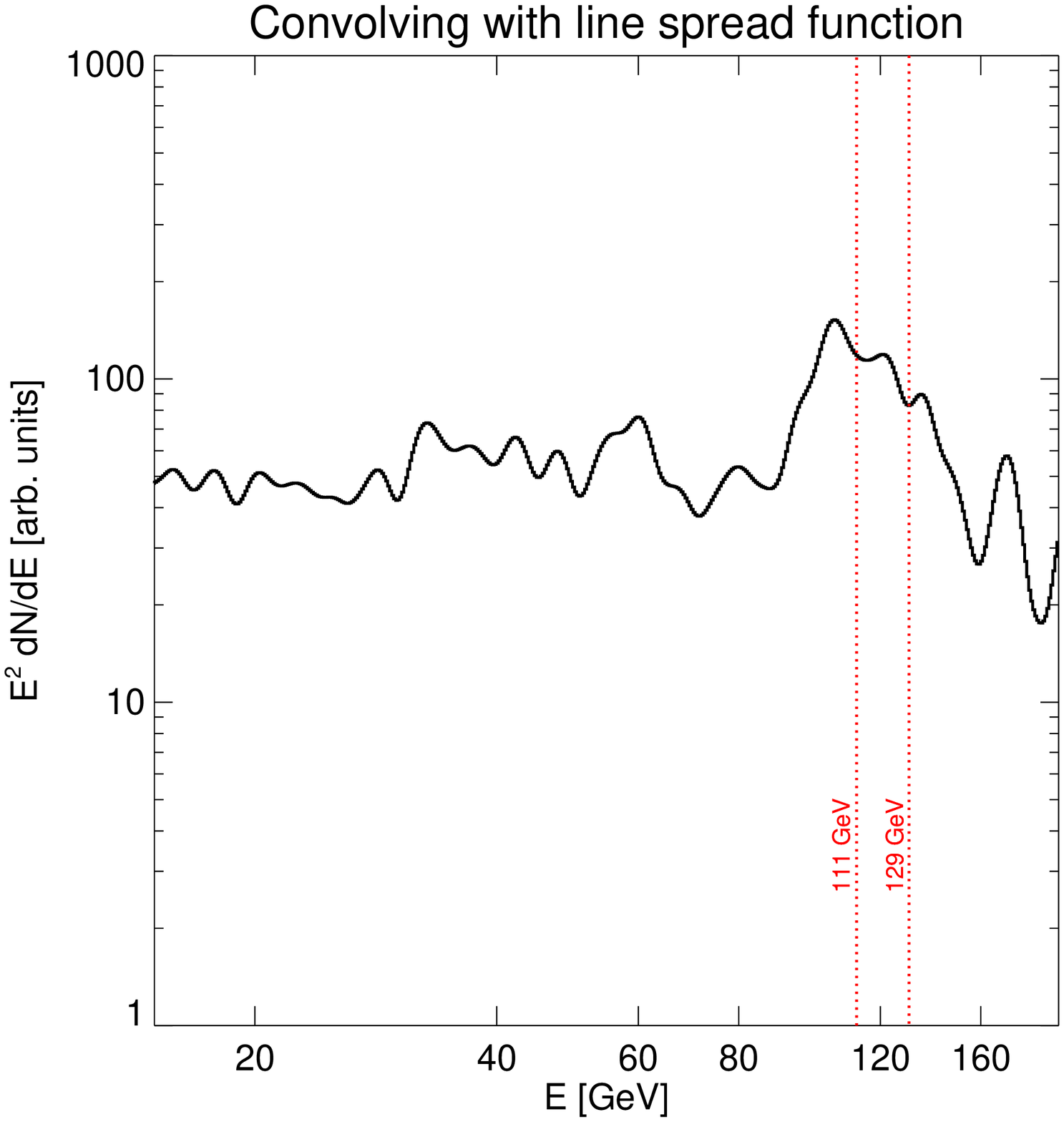}
  \end{center}
  \caption{Same as \reffig{fig1}, but using the \texttt{ULTRACLEAN} event
    class.}
  \label{fig:fig2}
\end{figure*}

\begin{figure}[ht]
  \begin{center}
    \includegraphics[width=0.45\textwidth]{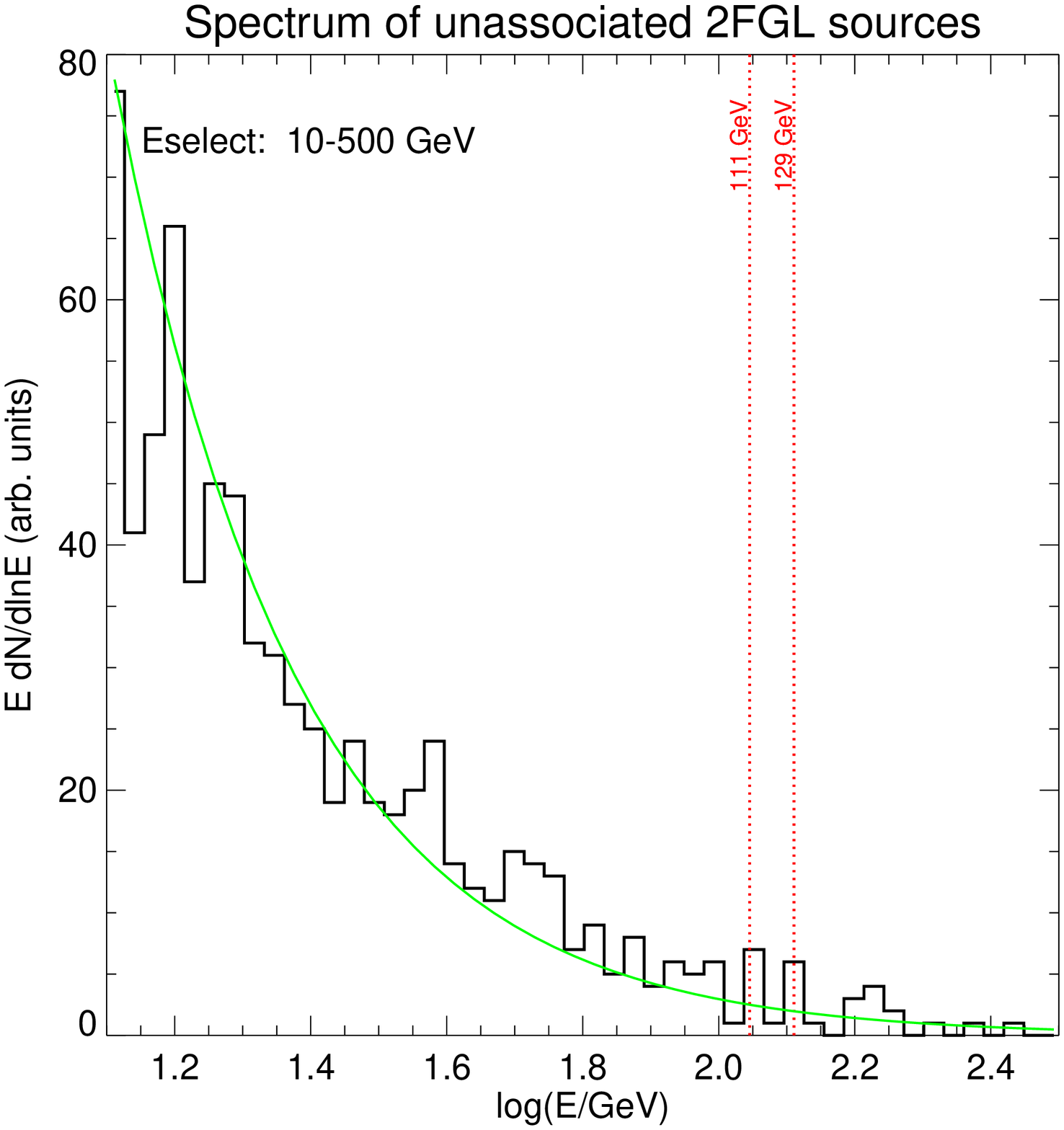}
    \includegraphics[width=0.45\textwidth]{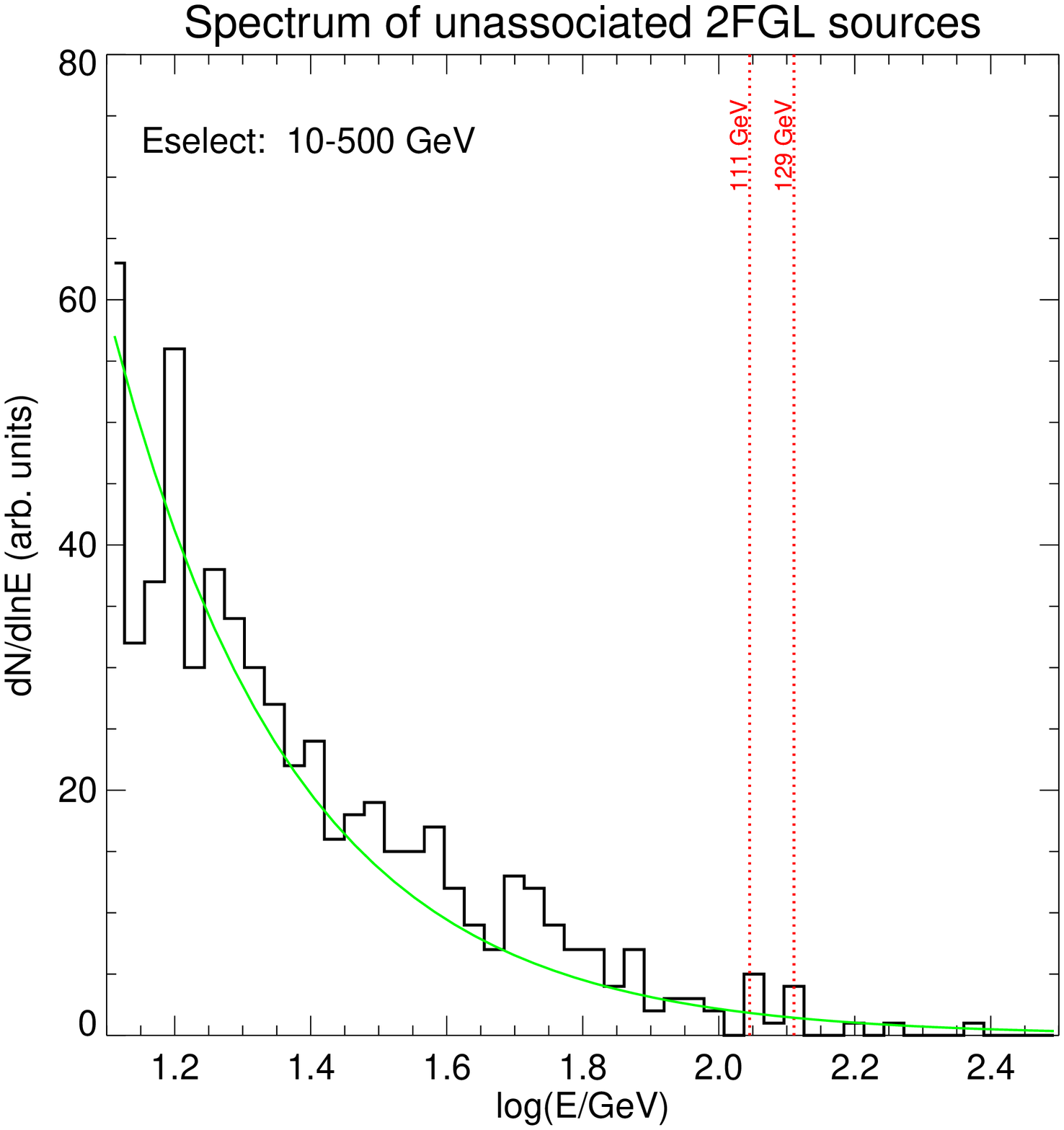}
  \end{center}
  \caption{Upper panel:  same as the upper left panel of \reffig{fig1} (with
    an extra power of $E$,
  but for all unassociated sources.  The green line is a power law
  $dN/dE\sim E^{-2.6}$ fit to events at $E>135$ GeV.  Lower panel:  same but
  for \texttt{ULTRACLEAN} class.  The suspicious bump at log$E$=2.2 in the
  upper panel disappears in the lower panel.  However, the background estimate
  in the lower panel appears to be unreasonably low, so the green line is
  multiplied by a factor of 2.5 to match the lower energy emission.  Because
  of this disagreement, we do not use the
  \texttt{ULTRACLEAN} data for our final result. }
  \label{fig:fig3}
\end{figure}

\begin{figure*}[ht]
    \begin{center}
        \includegraphics[width=0.45\textwidth]{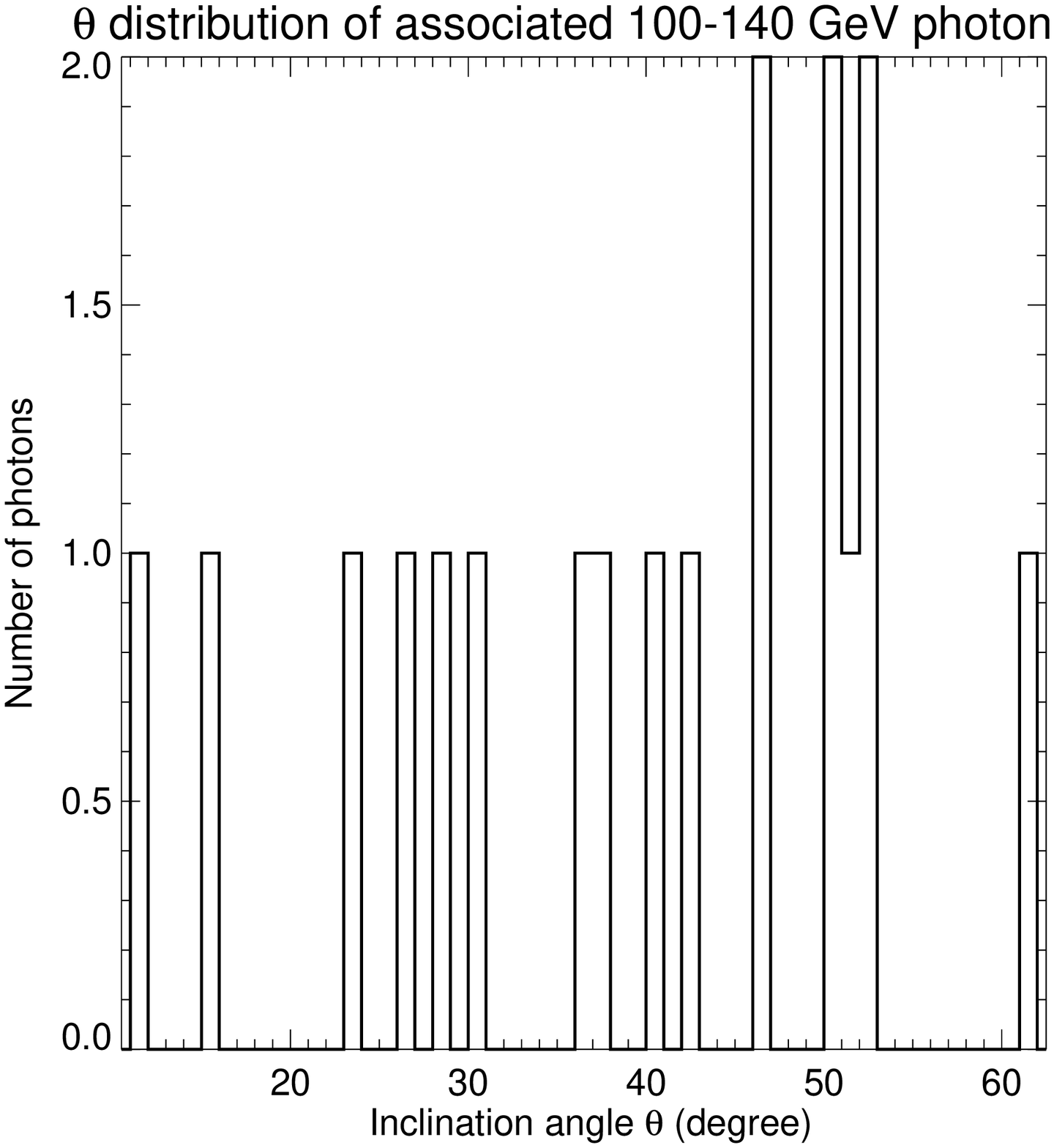}
        \includegraphics[width=0.45\textwidth]{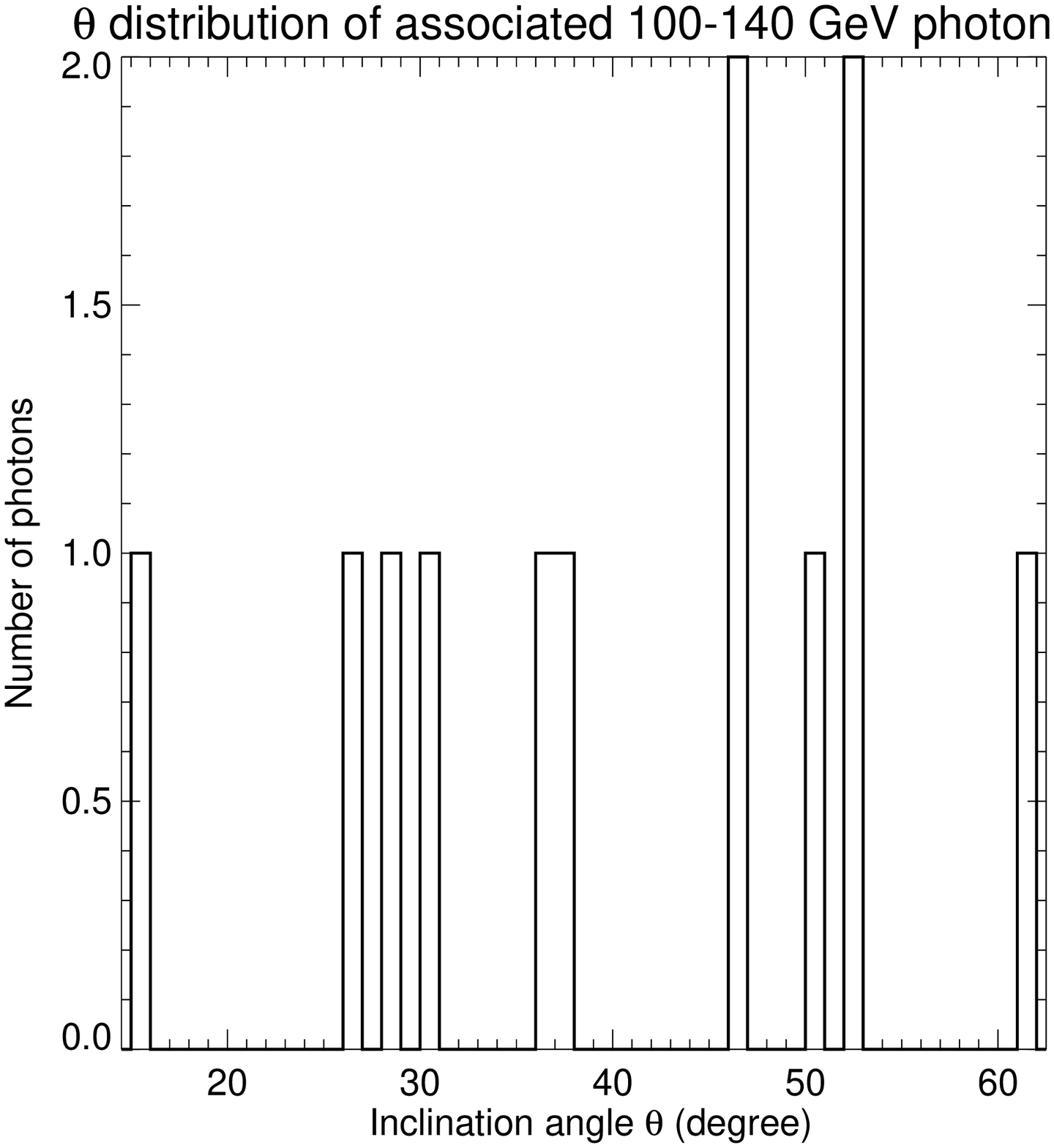}
    \end{center}
\caption{The inclination angle distribution of the 100-140
GeV photons in left panels of \reffig{fig1} and
\reffig{fig2}, i.e. for \texttt{SOURCE} and \texttt{ULTRACLEAN} events,
respectively. }
\label{fig:theta}
\end{figure*}


\begin{figure}[ht]
\begin{center}
\includegraphics[width=0.45\textwidth]{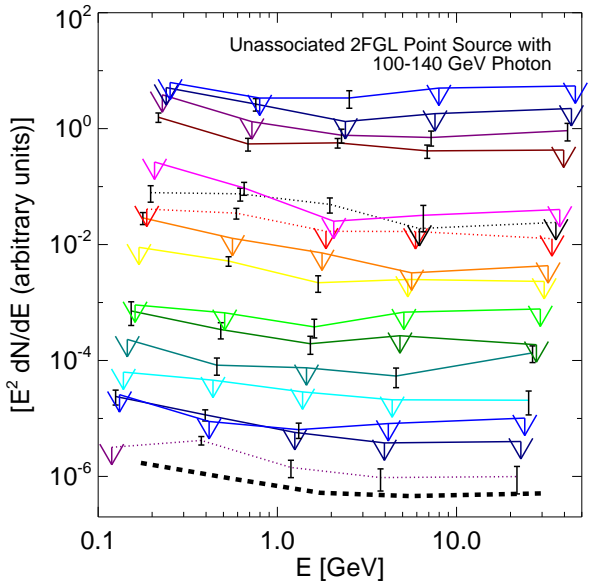}
\end{center}
\caption{The energy spectrum of unassociated 2FGL sources
which have a 100-140 GeV photon within
0.15$\degree$/0.3$\degree$ radius for \texttt{FRONT/BACK}
LAT events. The spectrum is obtained from the 2FGL
catalog. Sources thought to be potentially confused with Galactic
diffuse emission are shown with dotted lines. The dashed
black line shows the average of all the spectra. Each band
shows integral photon flux from [0.1-0.3, 0.3-1, 1-3, 3-10,
10-100] GeV respectively from the likelihood analysis in
that band with fixed photon power-law index.  A 2$\sigma$
upper limit is shown if the source is not significant in
a band.  }
\label{fig:fullspec}
\end{figure}

\begin{figure}[ht]
\begin{center}
\includegraphics[width=0.45\textwidth]{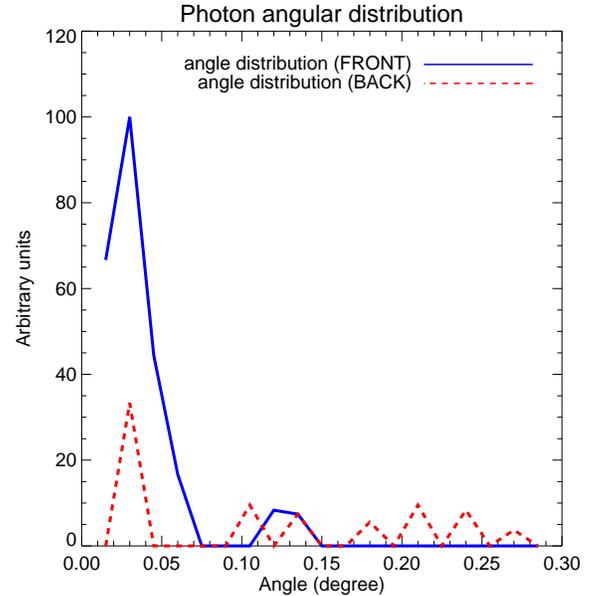}
\end{center}
\caption{The angular distribution of 100-140 GeV photons
matching with unassociated 2FGL sources within
0.15$\degree$/0.3$\degree$ radius for \texttt{FRONT/BACK}
LAT events (shown with solid/blue and dash/red curve,
respectively). The distribution is normalized by the annular
area given a radius. The \texttt{FRONT} events shows more
concentrated distribution than the \texttt{BACK} events,
consistent with the expectation based on the point spread
function. Both \texttt{FRONT} and \texttt{BACK} events
suggest a central concentrated distribution. }
\label{fig:profile}
\end{figure}


\begin{figure}[ht]
\begin{center}
\includegraphics[width=0.45\textwidth]{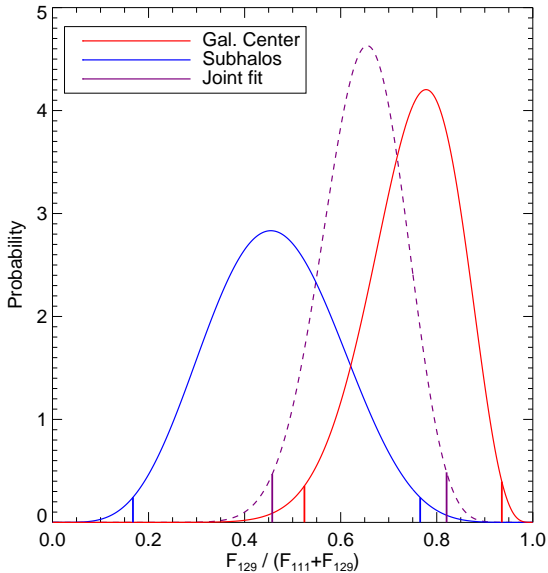}
\end{center}
\caption{Probability of obtaining the observed counts, in the energy bins
  centered on 111 and 129 GeV, in the Galactic center
  and subhalos as a function of the line fraction $f\equiv
  F_{129}/(F_{111}+F_{129})$. We find that the best fit ratio of the 129 GeV
  line to 111 GeV line is 1.5, and the 2$\sigma$ range of the line ratio is
  [0.84, 4.5].  See Section \ref{sec:lineratio} for details.}
\label{fig:lineratio}
\end{figure}


\begin{figure*}[ht]
\begin{center}
\includegraphics[width=0.8\textwidth]{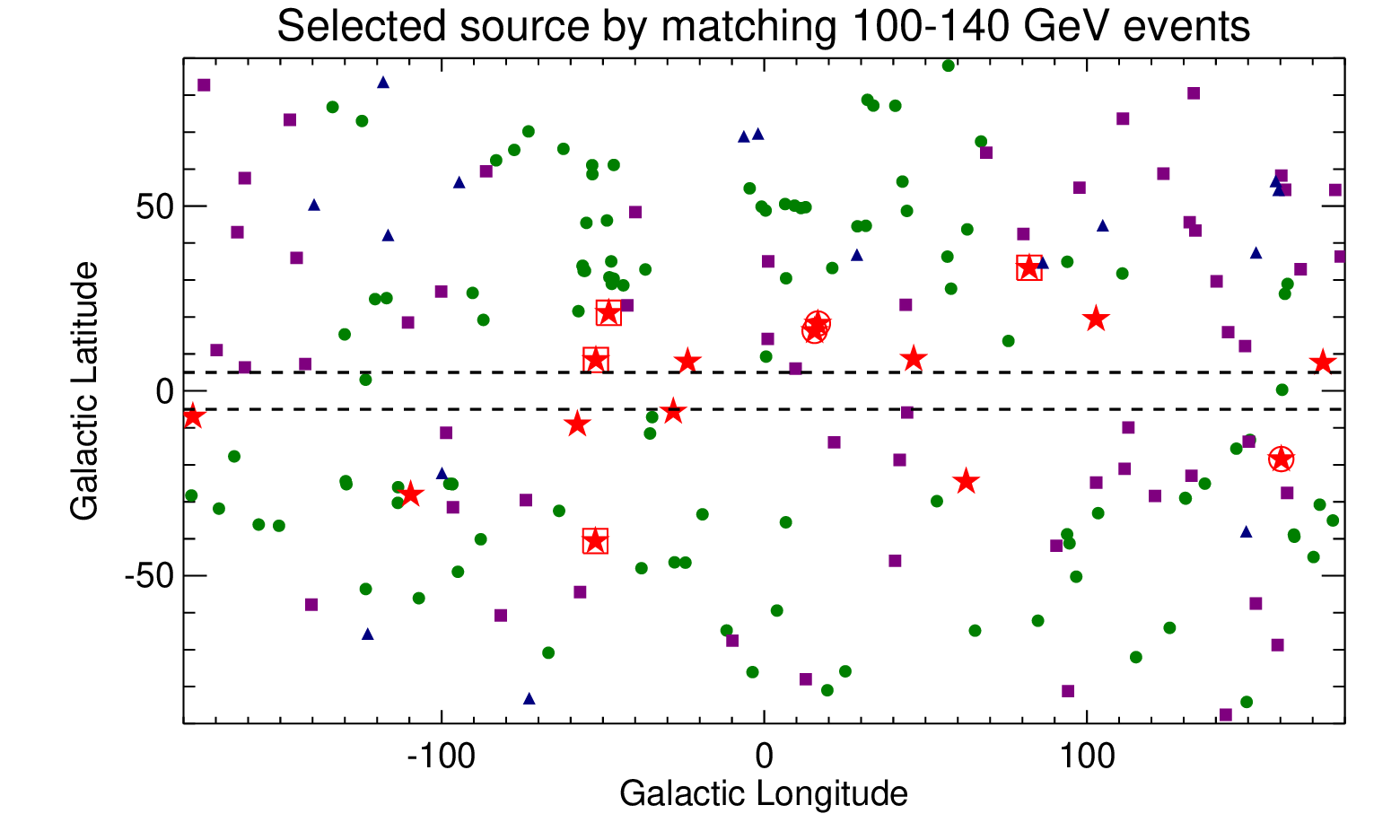}
\includegraphics[width=0.8\textwidth]{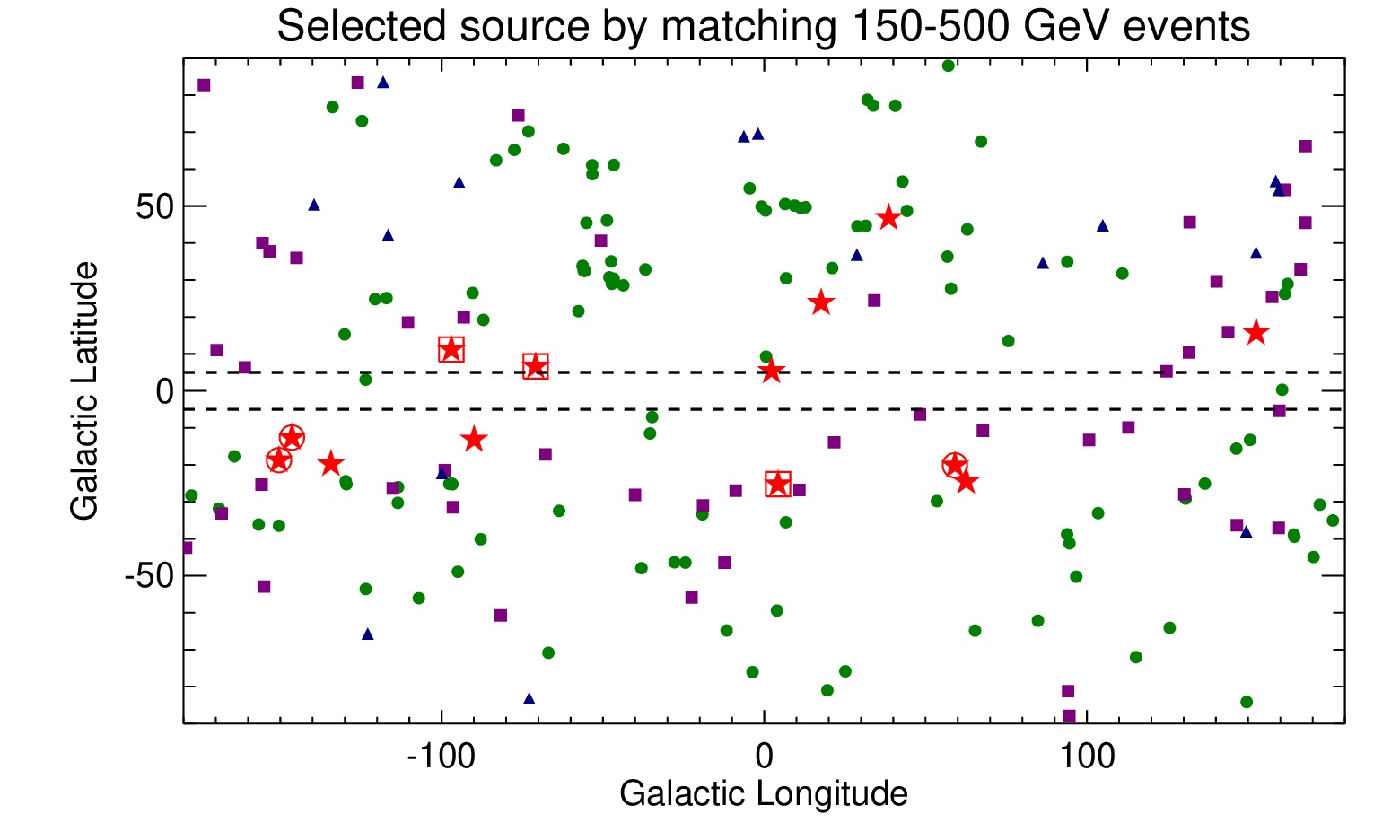}
\end{center}
\caption{The spatial distribution of unassociated 2FGL
sources which have a 100-140 GeV photon within
0.15$\degree$/0.3$\degree$ radius for \texttt{FRONT/BACK}
LAT events (red stars). For comparison, we also show the
spatial distribution of 16 dwarf galaxies
(e.g. \citealt{2010ApJ...712..147A}, blue triangles), 106 nearby
galaxy clusters (\citealt{2002ApJ...567..716R}, dark green
circles), and associated sources with the same matching
criteria (purple squares). We find no spatial overlaps
between these sources. Unassociated sources thought to be
potentially confused with Galactic diffuse emission (red
circles) and unassociated sources with less than 5$\sigma$
detection (red squares) are noted. The lower panel shows the
same plot as the upper panel but selected unassociated point
sources by matching with 150-500 GeV events.  In both cases,
the sources with high-energy photons are mostly near the
disk. This implies that the Galactic latitude distribution
may result from a selection effect.}
\label{fig:position}
\end{figure*}

In our recent paper~\citep{linepaper}, we have performed a
study with various data analysis methods and
obtained 6.5$\sigma$ local significance of the gamma-ray
study with various data analysis methods and
obtained 6.6$\sigma$ local significance of the gamma-ray
line structure, and 5.0-5.5$\sigma$ after trials factor
(depending on whether we assume one line or two lines). In
fact, we found {\em two} lines centered at 111 GeV and 129
GeV provide a better fit to the data.

The high significance of this result does not address
concerns about instrumental artifacts, such as energy
mapping errors that could give rise to spectral bumps and
dips \citep{syspaper}.  Such concerns must be addressed by
an independent analysis of photons from other parts of the
sky.  For example, detection of lines at 111 and 129 GeV
elsewhere on the sky in multiple unassociated LAT sources,
but \emph{not} in any class of associated LAT sources would
rule out an energy mapping error in the LAT data processing
as a source of the Galactic center lines.  Even a lower
significance detection would be interesting, because there is
no trials factor for choice of line energies.

Many of the \Fermi-LAT point sources are associated with counterparts at other
wavelengths, including blazars (BL Lacs, Flat Spectrum Radio Sources (FSRS),
etc.), other AGNs (Seyferts,
Radio Galaxies, etc.), pulsars and binaries, and other
Galactic sources~\citep{Fermi2FGL}. Although substantially
improved over the First {\it Fermi}-LAT catalog
\citep{Fermi1FGL}, there are still 344 sources in the 2FGL
($\sim$15\% of the total) without obvious counterparts at
Galactic latitude $|b| \geq 5^\circ$.

Various statistical methods and theoretical scenarios have
been suggested to classify and explain these unassociated
sources~\citep{FermiUnasso2012}, including the existence of
new types of source classes, e.g. dark matter subhalos.
Numerical simulations suggest
that within the Milky Way halo, dark matter subhalos form at all
mass scales down to the simulation resolution.
Less massive halos might show themselves as
gamma-ray sources without significant emission at other
wavelengths~\citep{2011arXiv1111.2613B,2012ApJ...747..121A,2010PhRvD..82f3501B}. If
such a signal were detected, it would be the first
non-gravitational signature of dark matter.

In this work, we use 3.9 years of LAT data to study the
gamma-ray spectrum of the unassociated point sources in the 2FGL
catalog to search for gamma-ray {\em line emission}. We
find that the energy spectrum shows two lines at 111 GeV
and 129 GeV. In \Refsec{data} we describe our
LAT data selection and analysis procedure. In \Refsec{spec}
we examine the spectral line emission by various
statistical tests and we summarize our main findings in
\Refsec{conclusion}.

\section{\emph{Fermi} data selection}
\label{sec:data}

In this section, we briefly describe our data selection and
analysis procedure. We refer to our previous papers for more
detailed information \citep{FermiBubble, FermiJet,
linepaper}. The \Fermi\ LAT is a pair-conversion telescope,
in which incoming photons convert to $e^+e^-$ pairs, which
are then tracked through the detector.  The arrival
direction and energy of each event are reconstructed, and
the time of arrival recorded. The LAT is designed to survey
the gamma-ray sky in the energy range from about 20 MeV to
several hundreds of GeV. The point spread function (PSF) is
about 0.8$\degree$ for 68\% containment at 1 GeV and
decreases with energy as $r_{68}\sim E^{-0.8}$, asymptoting
to $\sim$ 0.2$\degree$ at high energy.  It is convenient to distinguish
between front-converting and back-converting events that convert in the front
and back regions of the tracker, respectively.  The 68\% containment radius at
high energy is $r_{68}\sim 0.15\degree$ for front-converting and $r_{68}\sim
0.30\degree$ for back-converting events, with some dependence on the
incidence angle on the detector. 

The LAT energy resolution (i.e. the half-width of the 68\% containment region)
is of order $10$\% over most of the energy range \citep[see][for
details]{2012arXiv1206.1896F}.  Around 100 GeV, the resolution is closer to
7\% for high incidence-angle events, and twice that for normal incidence.

We use the latest publicly available data and instrument
response functions, known as Pass 7
(\texttt{P7\_V6})\footnote{Details at
\texttt{http://fermi.gsfc.nasa.gov/ssc/data/analysis/}
\texttt{documentation/Pass7\_usage.html}}. 
We perform our analysis on both \texttt{SOURCE} and
\texttt{ULTRACLEAN} events, and present figures based on each for comparison.
The former has larger effective area and higher background.  At the energies
of interest \texttt{CLEAN} and \texttt{ULTRACLEAN} events give identical
results. 


Photons coming from the bright limb at Earth's horizon, dominantly produced by
grazing-incidence CR showers in the atmosphere, are a potential source of
contamination. We remove this background by selecting events with zenith angle
less than $105\degree$.  We also exclude some time intervals when data quality
is poor, primarily while \Fermi\ passes through the South Atlantic Anomaly.

\section{Analysis}
\label{sec:spec}

High resolution simulations suggest the existence of
hierarchical dark matter halo structures on all resolved
mass
scales~\citep[e.g.][]{2009Sci...325..970K,2008MNRAS.384.1627P}.
The high end of the mass function is visible in the Milky
Way as dwarf galaxies, including the Magellanic clouds.
Less massive subhalos could be too small to contain enough
baryonic matter to be visible at other wavelengths, but
shine only via annihilation of dark matter particles in
gamma rays.  Such sources may appear as unassociated
\Fermi-LAT sources in the 2FGL catalog, which forms the
basis for our line search.

\subsection{Source selection}
The \Fermi-LAT 2FGL catalog consists of 1873 sources (100
MeV-100 GeV), of which 1290 are firmly identified/associated
and 575 (31\%) are unassociated sources \citep{Fermi2FGL}.
Cutting to Galactic latitude $|b| > 5\degree$ to avoid
contamination from the Galactic disk results in 344
unassociated sources. Since dark matter emission is expected
to be non-variable in time, we also remove the 25 among the
344 unassociated sources which have been flagged with
\texttt{variability\_index} $>$ 41.6. Then we select photon
events with energy in the range 100-140 GeV and zenith angle
less than 105$\degree$.  We include both \texttt{FRONT} and
\texttt{BACK} converting events. Among the remaining 319
unassociated sources we select ones with at least one
100-140 GeV photon within 0.15$\degree$/0.3$\degree$ for
\texttt{FRONT/BACK} converting events, which results in 16
unassociated sources for \texttt{SOURCE} event class. The
detailed information of these 16 sources can be found in
\reftbl{sourcelist}.

In \cite{Fermi2FGL}, it has been noted that 51\% of the unassociated sources
have been flagged due to various issues (compared to 14\% of the associated
sources).  We have tried cutting on each flag bit and find no significant
impact on our results.  In order to avoid introducing an additional trials
factor, we decided not to cut on any flags.

\subsection{Composite energy spectrum}
We show the energy spectrum of the regions within
0.15$\degree$/0.3$\degree$ for \texttt{FRONT/BACK}
converting events in two ways.  In the left panels of
\reffig{fig1} and \ref{fig:fig2}, we show the photon
counts binned in log $E$.  The binning used is the same as
in Fig. 16 of \cite{linepaper}.  This binning was chosen to
optimize the signal on the 111 GeV and 129 GeV lines in the
Galactic center, and has \emph{not} been modified for this
study.  In the right panels of \reffig{fig1}, we plot $E^2
dN/dE$ for the unbinned event distribution convolved with
the line-spread function (LSF)~\citep{Edmonds:thesis,
  linepaper}.  Each of these representations has pros and
cons.  The histograms make it easy to see how many photons
contribute to each bin, and allow simple computations of
Poisson likelihoods.  The unbinned spectra give a sense of
the spectrum without any arbitrary binning choices, and
convolution by the LSF allows maximum sensitivity to faint
signals.  However, the spectrum is ``twice convolved'' (once
by the instrument and once by the processing) making the
lines blend together to an undesirable degree.  In the
following we use the binned histograms for analysis, and
provide the smoothed spectra only for reference.

We plot the energy spectrum for photons near the 16 unassociated sources
with a photon at 100-140 GeV for \texttt{SOURCE} (\reffig{fig1}) and
\texttt{ULTRACLEAN} (\reffig{fig2}) event classes. 
While this selection obviously suppresses
the spectrum outside of the 100-140 GeV range, there is no way it can
rearrange photons in the 100-140 GeV window.  Intriguingly, we find two
gamma-ray emission lines at 111 GeV and 129 GeV.  One interpretation is that
unassociated sources emit a gamma-ray doublet.  Another is that
some flaw in the LAT data preferentially maps events to these energies.  In
order to test this hypothesis, we apply the {\em same} selection procedure to
associated point sources, also shown in Figures \ref{fig:fig1} and
\ref{fig:fig2}, and find no line features at 111 GeV or 129 GeV.

For each of the 16 unassociated sources, we show the
integrated photon flux in energy bands (0.1-0.3, 0.3-1, 1-3,
3-10, 10-100) GeV respectively in \reffig{fullspec},
obtained from the 2FGL catalog. Among the 16 sources, three
of them are marked with potentially confused with Galactic
diffuse emission. 

\subsection{Statistical significance}
Even with low statistics (7 counts at 111 GeV and 6 counts at 129 GeV) it is
possible to obtain a significant result if the backgrounds are low enough.
Because WIMP annihilations can produce lower energy photons (final-state
radiation, Z/W continuum, inverse-Compton, etc.) it may be incorrect to use
lower energy emission to assess the background.  However, at high energy there
are very few photons in these sources, and there would be none from a 129 GeV
WIMP.  As a compromise, we assume the background is a power law, fit its
amplitude to high energy ($135 < E < 270$), but choose the power-law index so
that lower energy emission is modeled approximately correctly (\reffig{fig3}).

We assess the Poisson probability of observing 13 (or more) \texttt{SOURCE}
counts in the two spectral bins with the background estimate in the upper
panel of \reffig{fig3}.  This has a probability of $p=0.00069$ corresponding
to 3.2$\sigma$. Removing sources to be potentially confused with Galactic
diffuse emission (marked out in 2FGL) only mildly affect our results
(3.3$\sigma$). The \texttt{ULTRACLEAN} events would give a much higher
significance ($>4\sigma$) if we could believe the background estimate, but it
looks implausibly low.

\subsection{Spatial distribution}
The subhalo candidates identified in this work are mostly distributed at $|b|
< 20\degree$, at all longitudes.  It is not clear whether this could be a
selection effect, a fluke, or a hint about the true distribution of dark
matter subhalos.  On one hand, dark matter subhalos preferentially dragged
into the Galactic disk may lead to disk-like configurations, e.g. the proposed
``dark disk'' \citep[e.g.][]{Bruch:2009,Purcell:2009}. On the other hand, the
distribution is not concentrated in longitude, so they may be nearby subhalos
with lower mass, close enough to appear brighter than more massive subhalos,
e.g. those hosting dwarf galaxies.

\subsection{Radial profile}
In \reffig{profile}, we show the stacked angular
distribution of 100-140 GeV photons, which are selected by
matching with unassociated 2FGL sources within
0.15$\degree$/0.3$\degree$ radius for \texttt{FRONT/BACK}
LAT events, with respect to the source center provided by
2FGL. The distribution is normalized by the annular area
at each radius. The \texttt{FRONT} events show a more
concentrated distribution than the \texttt{BACK} events,
consistent with the point spread
function. Both \texttt{FRONT} and \texttt{BACK} events
suggest a centrally concentrated distribution.

\subsection{Line ratio}
\label{sec:lineratio}
Our previous work \citep{linepaper} found 4 (14) photons above background at
111 (129) GeV.  This led us to expect the 129 GeV line might be stronger, but
this work finds the 111 GeV to have slightly more counts:
6 (5) at 111 (129) GeV above background.  Are these results compatible?

In order to determine a confidence interval for the line ratio, we consider a
total of $N$ photons for the doublet, with $k$ of them in the 129 GeV bin, and
the rest in the 111 GeV bin.  The binomial probability of observing $k$ of $N$
counts in this bin is
\be
P_b(k,n,f) = \frac{N!}{k!(N-k)!} f^k (1-f)^{N-k}
\ee
where $f\equiv F_{129}/(F_{111}+F_{129})$ is the true fraction of doublet
photons at 129 GeV.  Figure \ref{fig:lineratio} shows this probability (i.e.,
the probability of observing $k$ counts given $N$ and $f$) as a function of
$f$ for the GC, subhalos, and the product of the two.

To obtain a confidence interval, we find $f_{\rm low}$ such that 
\be
P(k\ge x,n,f_{\rm low}) = \sum_{k=x}^{N} P_b(k,n,f) = 0.025
\ee
with a complementary expression for $f_{\rm high}$.  The 95\% confidence
interval (corresponding to ``2$\sigma$'' confidence) is then 
$0.167 < f < 0.765$ for the subhalos and 
$0.524 < f < 0.935$ for the Galactic center.  A significant range of $f$ is
allowed in both cases, so we can combine the counts from both and obtain
$0.457 < f < 0.820$ for the joint fit.
This yields 95\% confidence bounds on the line ratio $0.84 <
F_{129}/F_{111} < 4.5$.  The data are consistent (at $2\sigma$) with the lines
being equally strong, but also with the 129 GeV line being 4.5 times as
strong.  Clearly more data will be required to measure the line ratio with
high confidence. 

\section{Discussion and Conclusion}
\label{sec:conclusion}

In this paper, we have reported evidence for line emission at 111 GeV and 129
GeV from unassociated \Fermi-LAT point sources.  The lines have a significance
of $p=6.9\times10^{-4}$ or $3.2\sigma$ for a simple power-law background
model.  These results provide independent support for our previous claims of a
double gamma-ray line in the Galactic center at the {\em same}
energies~\citep{linepaper}. The double line emission is compatible with the
scenario of a 129 GeV WIMP annihilating to $\gamma \gamma$ and $\gamma Z$,
producing the two lines.  As a test of systematics, we apply the same
selection and analysis procedure to associated \Fermi-LAT point sources, and
find no evidence for lines at these energies. It is difficult to imagine
instrumental systematics that could produce this double line emission only at
the Galactic center region and the locations of unassociated point sources
without affecting other regions of the sky.  We find this evidence to be
persuasive, but it cannot be considered conclusive until more data become
available.

Further observations are essential, not only to firmly establish the existence
of the lines, but to measure the line ratio.  The ratio of $\gamma Z$ and
$\gamma \gamma$ line strength depends only on the particle physics model, and
is independent of the astrophysical uncertainties such as the dark matter
distribution in the halo.  The line pair is also compatible with a 141 GeV
WIMP annihilating through $\gamma Z$ and $\gamma h$ for $m_h\sim125$ GeV, as
in the ``Higgs in Space'' scenario~\citep{Jackson:2010}.  However, we have not
found any significant gamma-ray line at $\sim 141$ GeV.  In any case, additional
data will be critical for measuring the line ratio, which is currently only
poorly determined (\reffig{lineratio}).

A possible change to the \Fermi\ scan strategy could accumulate S/N on the
double spectral lines in the Galactic center up to $\sim$4 times as fast as
the current survey strategy, and it is crucial for studying the double line
emission.  We believe this could be done with only modest impact on other
\Fermi\ science objectives.  With the huge effective area and low energy
threshold of H.E.S.S II, it may be possible to confirm a spectral bump in the
Galactic center fairly soon.  However, the energy resolution of H.E.S.S. II is
inferior to LAT at 129 GeV, and it may be difficult to resolve the doublet. 

The 2FGL catalog is based on 24 months of LAT data, and an updated catalog
based on 48 months would provide improved source parameters and associations,
decreasing the background noise for the subhalo analysis presented here.
Furthermore, multi-wavelength follow-up observations would be helpful to
identify the nature of unassociated 2FGL sources and refine the list of
associations. 

By stacking the unassociated 2FGL point sources together, it might be possible
to reveal the spatial profile of the gamma-ray distribution at 111 GeV and 129
GeV. One may in principle improve the positional information of these
unassociated sources by using high energy photons and better quantifying the
gamma-ray spatial profile.  However, with $\sim 1$ photon per source at high
energy, the details of the algorithm used for centroiding the sources are
critical, and consideration of the spatial profile is beyond the scope of this
work.  We simply use the position provided by the \Fermi\ 2FGL catalog.


\begin{deluxetable*}{||l|r|r|r|r|r|r|r|r|r|r|r||}
  \tablehead{ Source name & RA & Dec & $\ell$ & b & Flags & $\sigma$ & Variability & Spec index & Radius & Spec type}
  \startdata
2FGL J0341.8+3148c
 &   55.5 &   31.8 &  160.3 &  -18.4 &   32 &    7.6 &   24.7 &   2.31 $\pm$  0.09 & 0.0706 & PowerLaw           \\
2FGL J0526.6+2248 
 &   81.7 &   22.8 &  182.9 &   -6.9 &    1 &    7.1 &   27.5 &   2.88 $\pm$  0.16 & 0.0641 & PowerLaw           \\
2FGL J0555.9-4348 
 &   89.0 &  -43.8 &  250.4 &  -28.0 &    0 &    5.0 &   29.1 &   2.11 $\pm$  0.17 & 0.0994 & PowerLaw           \\
2FGL J0600.9+3839 
 &   90.2 &   38.7 &  173.2 &    7.6 &    0 &    5.1 &   14.1 &   2.04 $\pm$  0.21 & 0.0490 & PowerLaw           \\
2FGL J1240.6-7151 
 &  190.2 &  -71.9 &  302.1 &   -9.0 &    0 &    8.2 &   18.4 &   1.82 $\pm$  0.16 & 0.0458 & PowerLaw           \\
2FGL J1324.4-5411 
 &  201.1 &  -54.2 &  307.8 &    8.4 &   24 &    4.9 &   23.3 &   2.39 $\pm$  0.14 & 0.1126 & PowerLaw           \\
2FGL J1335.3-4058 
 &  203.8 &  -41.0 &  311.8 &   21.1 &    0 &    4.6 &   11.2 &   2.13 $\pm$  0.18 & 0.0800 & PowerLaw           \\
2FGL J1601.1-4220 
 &  240.3 &  -42.3 &  336.3 &    7.9 &    0 &    7.3 &   20.8 &   2.46 $\pm$  0.10 & 0.1037 & PowerLaw           \\
2FGL J1639.7-5504 
 &  249.9 &  -55.1 &  331.8 &   -5.6 &    9 &    5.9 &   21.1 &   2.79 $\pm$  0.14 & 0.0568 & PowerLaw           \\
2FGL J1716.6-0526c
 &  259.2 &   -5.4 &   16.6 &   18.2 & 2080 &    6.8 &   25.7 &   2.43 $\pm$  0.26 & 0.1052 & LogParabola        \\
2FGL J1721.5-0718c
 &  260.4 &   -7.3 &   15.6 &   16.2 &   41 &    6.0 &   20.9 &   2.68 $\pm$  0.36 & 0.0795 & LogParabola        \\
2FGL J1730.8+5427 
 &  262.7 &   54.5 &   82.2 &   33.3 &    0 &    4.6 &   16.9 &   2.69 $\pm$  0.18 & 0.1258 & PowerLaw           \\
2FGL J1844.3+1548 
 &  281.1 &   15.8 &   46.3 &    8.7 &    4 &   12.5 &   29.8 &   2.43 $\pm$  0.08 & 0.0403 & PowerLaw           \\
2FGL J2004.6+7004 
 &  301.2 &   70.1 &  102.9 &   19.5 &    0 &    9.5 &   36.8 &   1.97 $\pm$  0.11 & 0.0368 & PowerLaw           \\
2FGL J2115.4+1213 
 &  318.9 &   12.2 &   62.6 &  -24.5 &    0 &    5.1 &   25.3 &   2.38 $\pm$  0.19 & 0.0800 & PowerLaw           \\
2FGL J2351.6-7558 
 &  357.9 &  -76.0 &  307.7 &  -40.6 &    0 &    4.1 &   20.8 &   1.92 $\pm$  0.19 & 0.0702 & PowerLaw  
  \enddata
  \tablecomments{The table provides detailed information about the
unassociated 2FGL point sources we have identified with at
least one 100-140 GeV photon within 0.15/0.3$\degree$
for \texttt{FRONT/BACK} events. The first column is the 2FGL catalog name in
the format \texttt{2FGL
JHHMM.m+DDMM}, where 'c' indicates that the source
is considered to be potentially confused with Galactic
diffuse emission. The second/third column are Right
Ascension (J2000) and Declination (J2000). The forth and
fifth column are Galactic Longitude and Galactic
Latitude. The sixth column is the flag parameter which
indicate possible issues noted in detection or
characterization of the source. Sources having no flags
raised with value 0 are those without potential problems.
The seventh column is the variability index, defined as the
test statistic for the hypothesis that monthly averages of the source flux
vary relative to the null hypothesis of constant flux.  The TS is distributed
as $\chi^2$ with 23 degrees of freedom, so a
value greater than 41.64 indicates a $> 99$\% chance of being a
variable source, and we have removed these sources from our analysis. The
eighth column shows the best fit for the photon number
power-law index (for logarithmic parabola spectra it is index at the
Pivot Energy) derived from the likelihood analysis for 100
MeV-100 GeV.  The ninth column shows the average of
semimajor/semiminor axis of the error ellipse at 68\%
confidence. Source detection significance in Gaussian
$\sigma$ units is shown in the tenth column, which is
derived from the likelihood Test Statistic for 100 MeV-100
GeV analysis. The eleventh column shows the best fit form of
the spectral type. We note that only three sources have
logarithmic parabolic spectral shape (two of them are marked
with potentially confused with Galactic diffuse emission). 
Detailed explanation of parameters listed in this table can
be found in \cite{Fermi2FGL}.  }
\label{tbl:sourcelist}
\end{deluxetable*}



\vskip 0.15in {\bf \noindent Acknowledgments:} We thank
Christoph Weniger, Dan Hooper, and Neal Weiner for helpful
conversations.  We acknowledge the use of public data from
the \Fermi\ data archive at
\texttt{http://fermi.gsfc.nasa.gov/ssc/}.  This work would
not be possible without the work of hundreds of people, over
many years, to design, build, and operate \Fermi.  M.S. and
D.P.F. are partially supported by the NASA Fermi Guest
Investigator Program.  This research made use of the NASA
Astrophysics Data System (ADS) and the IDL Astronomy User's
Library at Goddard (Available at
\texttt{http://idlastro.gsfc.nasa.gov}).
\bibliography{unassociated_psc}
\bibliographystyle{hapj}

\appendix

\begin{figure*}[ht]
  \begin{center}
    \includegraphics[width=0.45\textwidth]{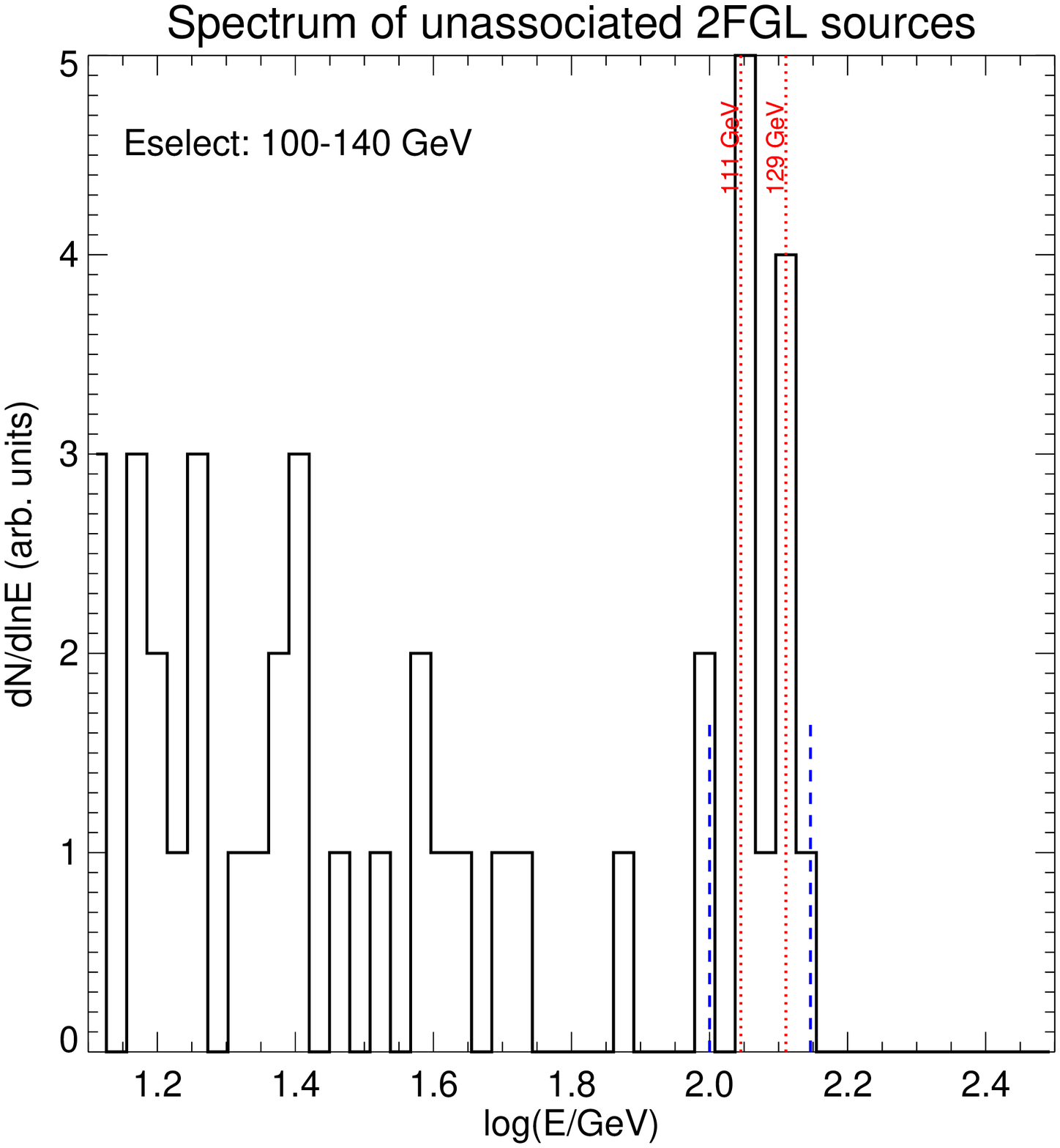}
    \includegraphics[width=0.45\textwidth]{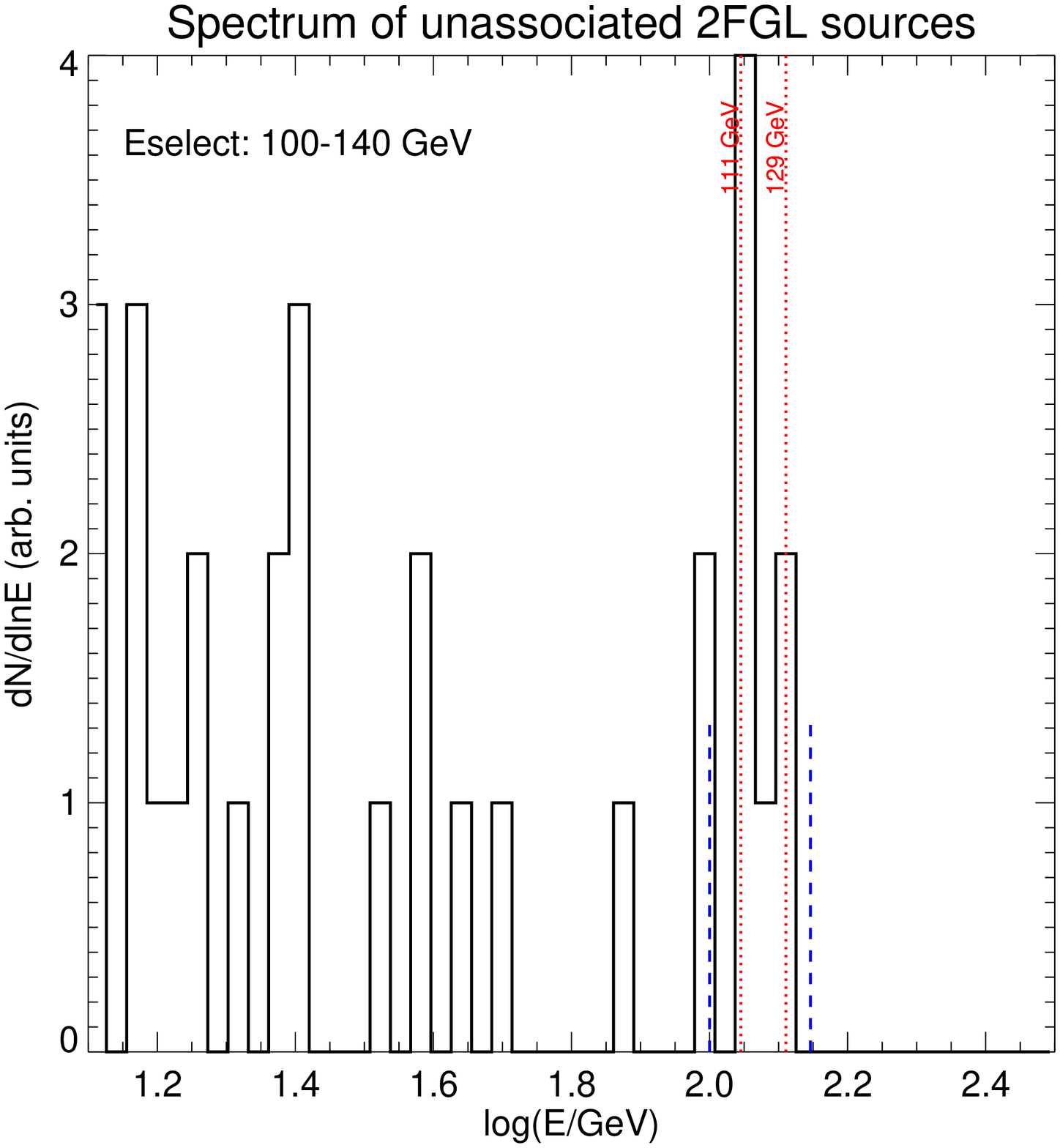}
  \end{center}
  \caption{Same as the upper left panel of \reffig{fig1},
but using only events with high incidence angle $\theta >
40\degree$ which has better energy resolution to reveal the
energy spectrum of unassociated 2FGL catalog. We found two
gamma-ray line emission on 111 GeV and 129 GeV. 
  }
  \label{fig:figapp1}
\end{figure*}

\begin{figure}[ht]
  \begin{center}
    \includegraphics[width=0.45\textwidth]{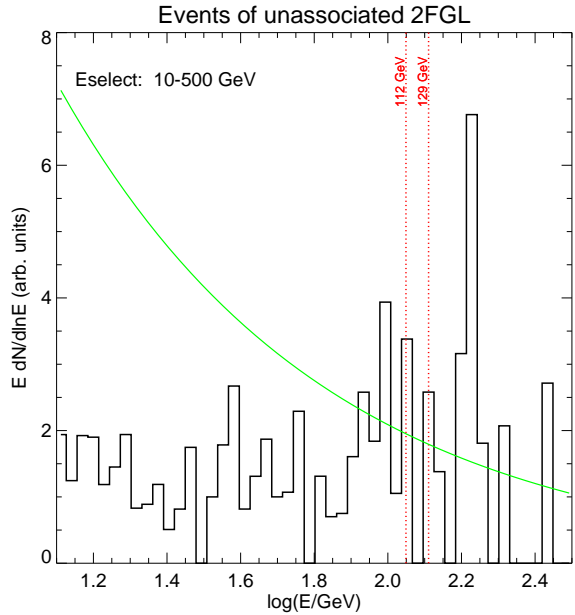}
  \end{center}
  \caption{Same as the upper left panel of \reffig{fig1},
but using only events belongs to \texttt{SOURCE} class but
{\em not} to \texttt{ULTRACLEAN} event. This selected set of
events are dominated by cosmic ray events. We found no
significant gamma-ray line emission on 111 GeV and 129 GeV
($0.47\sigma$ with background estimation using 100-500 GeV
data).  }
  \label{fig:figapp2}
\end{figure}

\begin{figure}[ht]
  \begin{center}
    \includegraphics[width=0.3\textwidth]{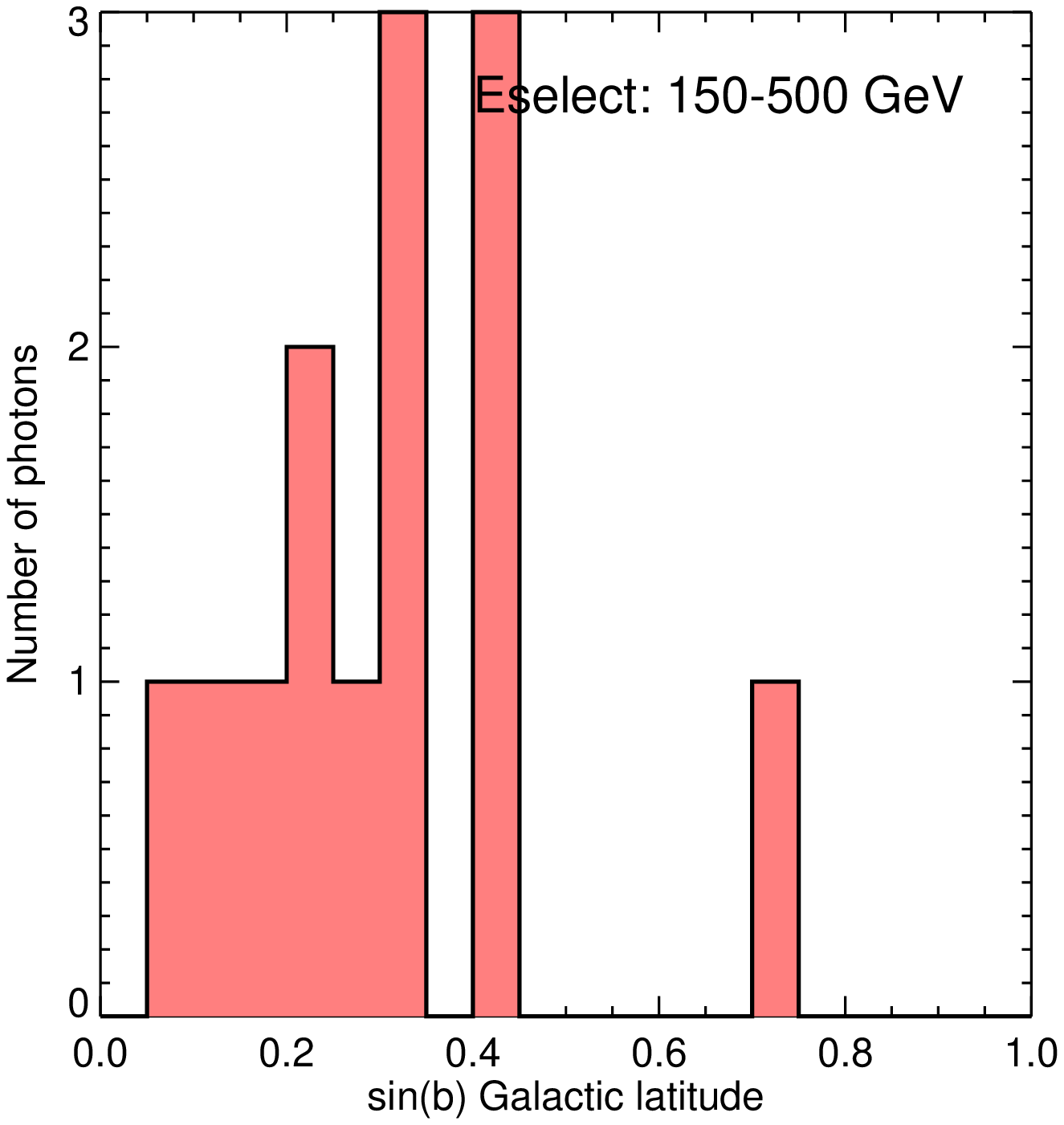}
    \includegraphics[width=0.3\textwidth]{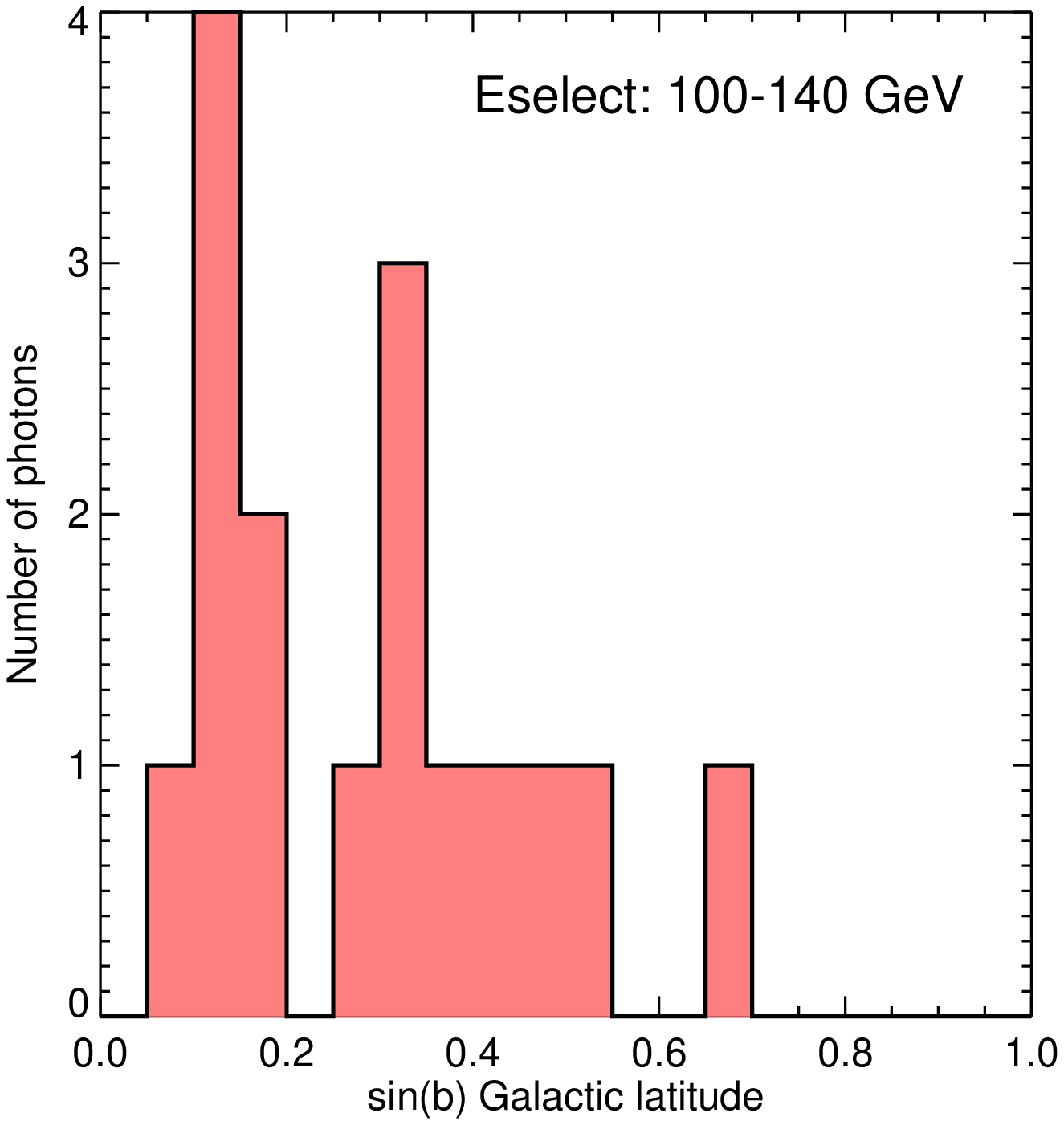}
    \includegraphics[width=0.3\textwidth]{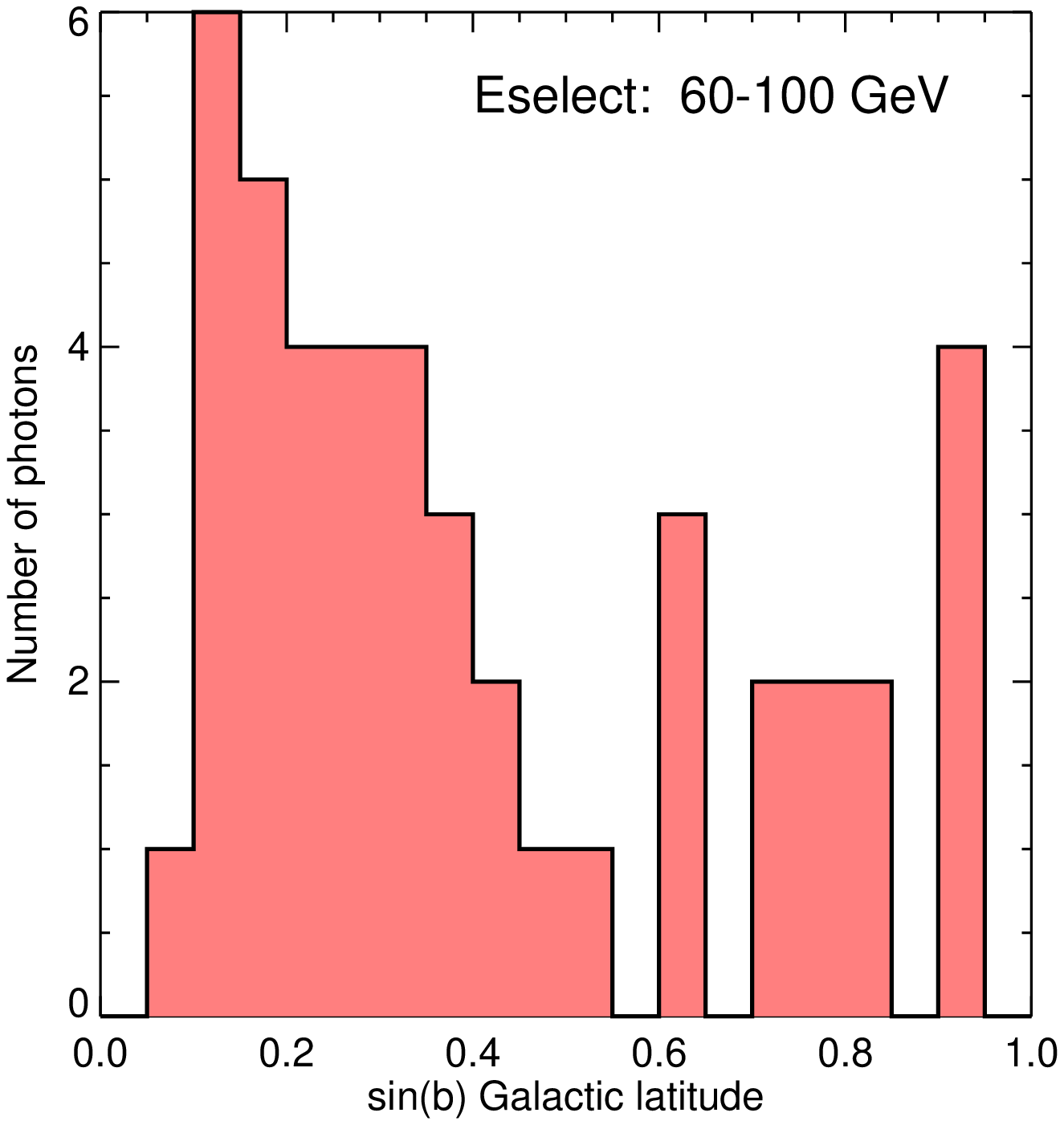}
    \includegraphics[width=0.3\textwidth]{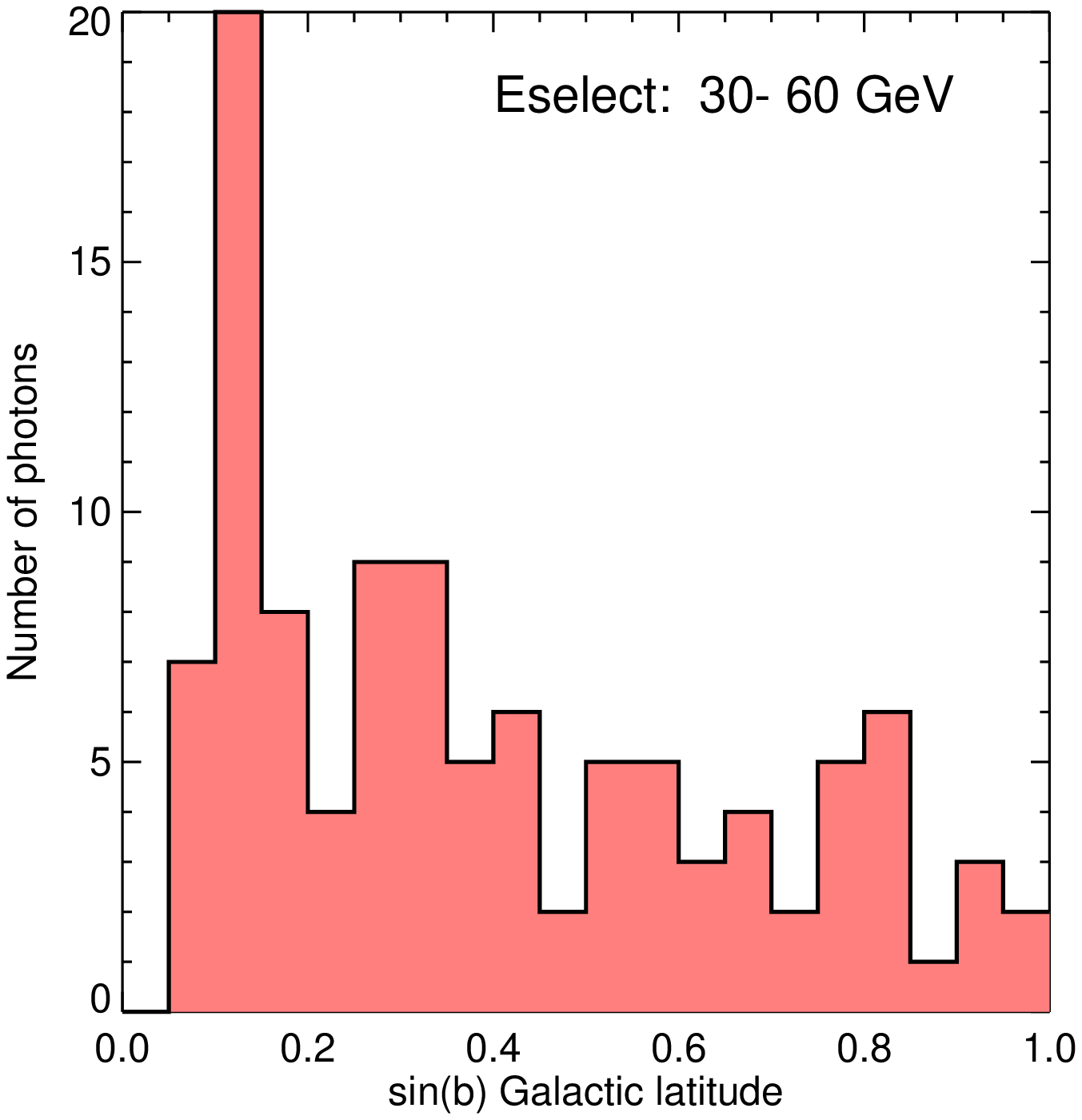}
    \includegraphics[width=0.3\textwidth]{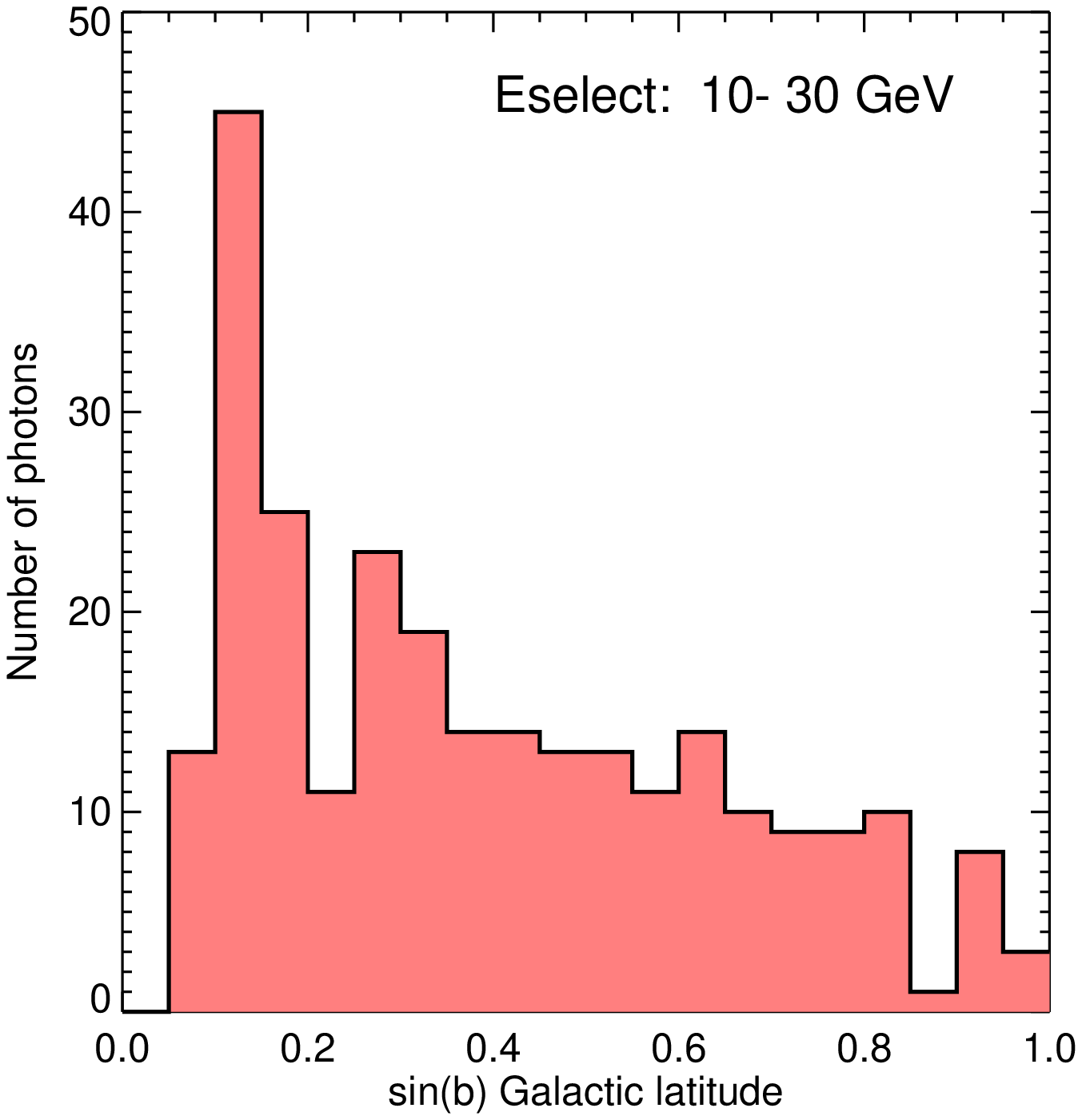}
    \includegraphics[width=0.3\textwidth]{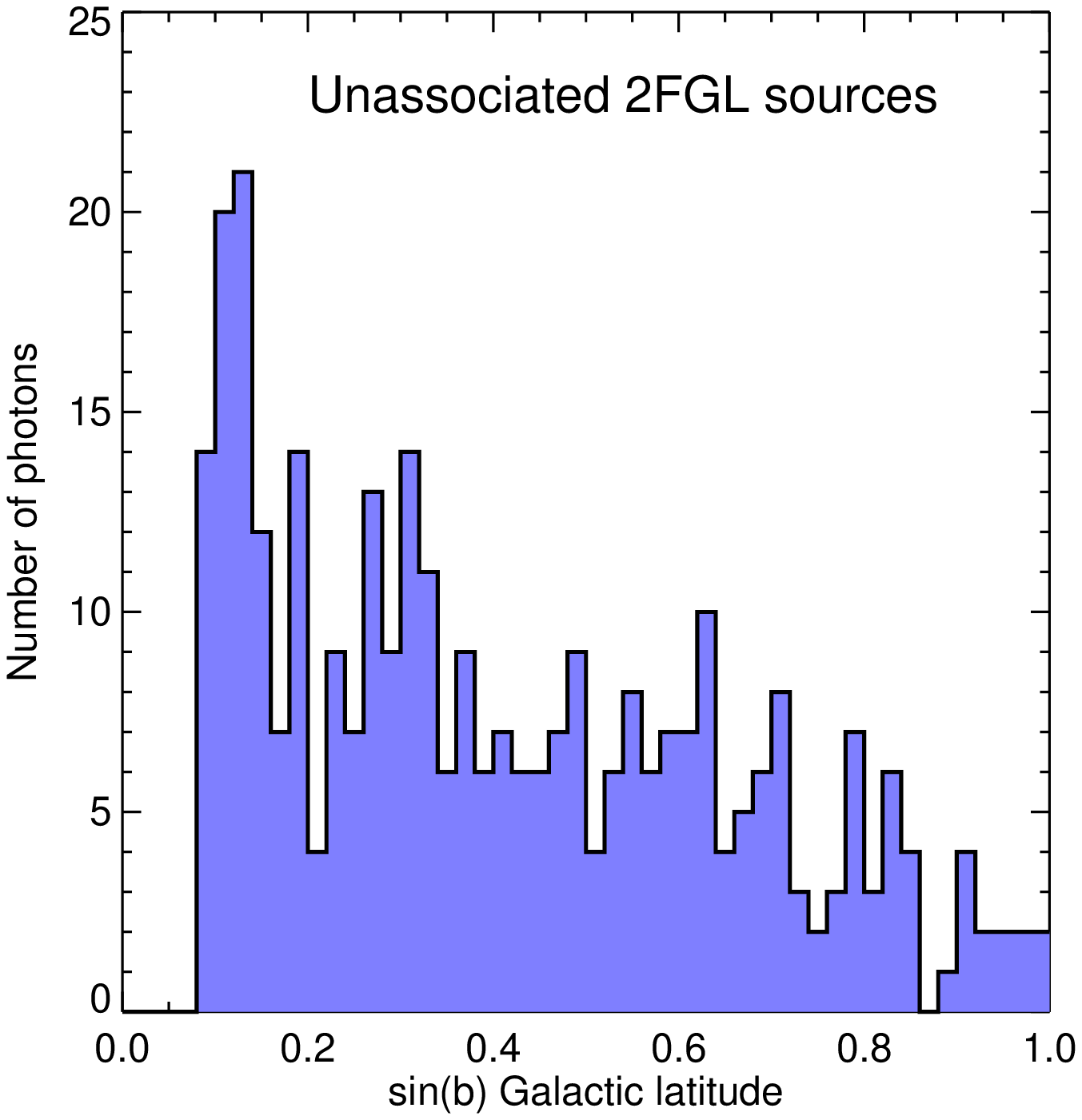}
  \end{center}
  \caption{The distribution of unassociated 2FGL sources
  selected by matching with LAT photons of different energy
  range. The panels from the upper left to lower right are
  for 150-500 GeV, 100-140 GeV, 60-100 GeV, 30-60 GeV, and
  10-30 GeV, and for comparison all unassociated sources
  with $|b| > 5\degree$ and variability index $<$ 41.64.
  }
  \label{fig:figapp3}
\end{figure}

\begin{figure}[ht]
  \begin{center}
    \includegraphics[width=0.3\textwidth]{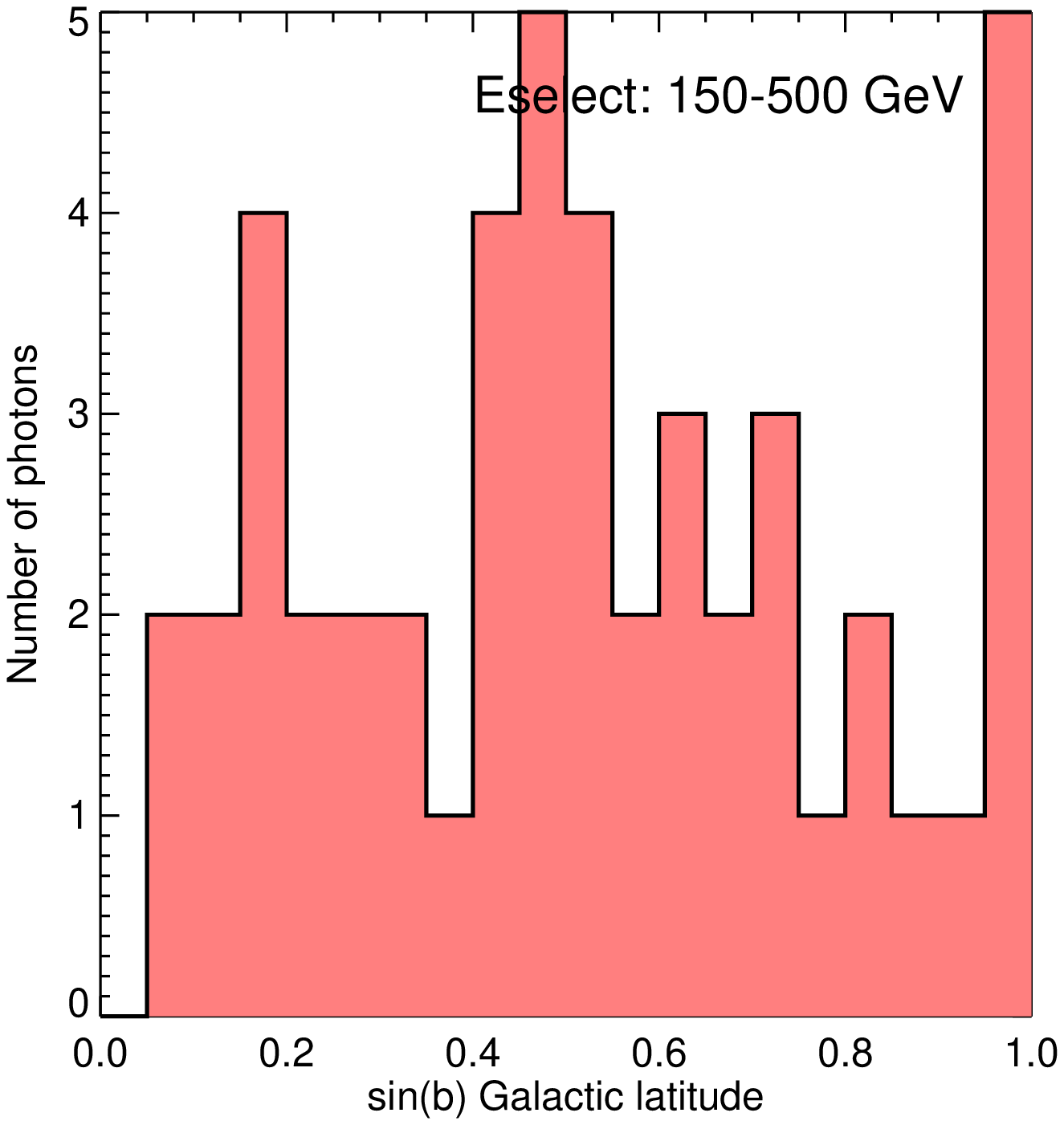}
    \includegraphics[width=0.3\textwidth]{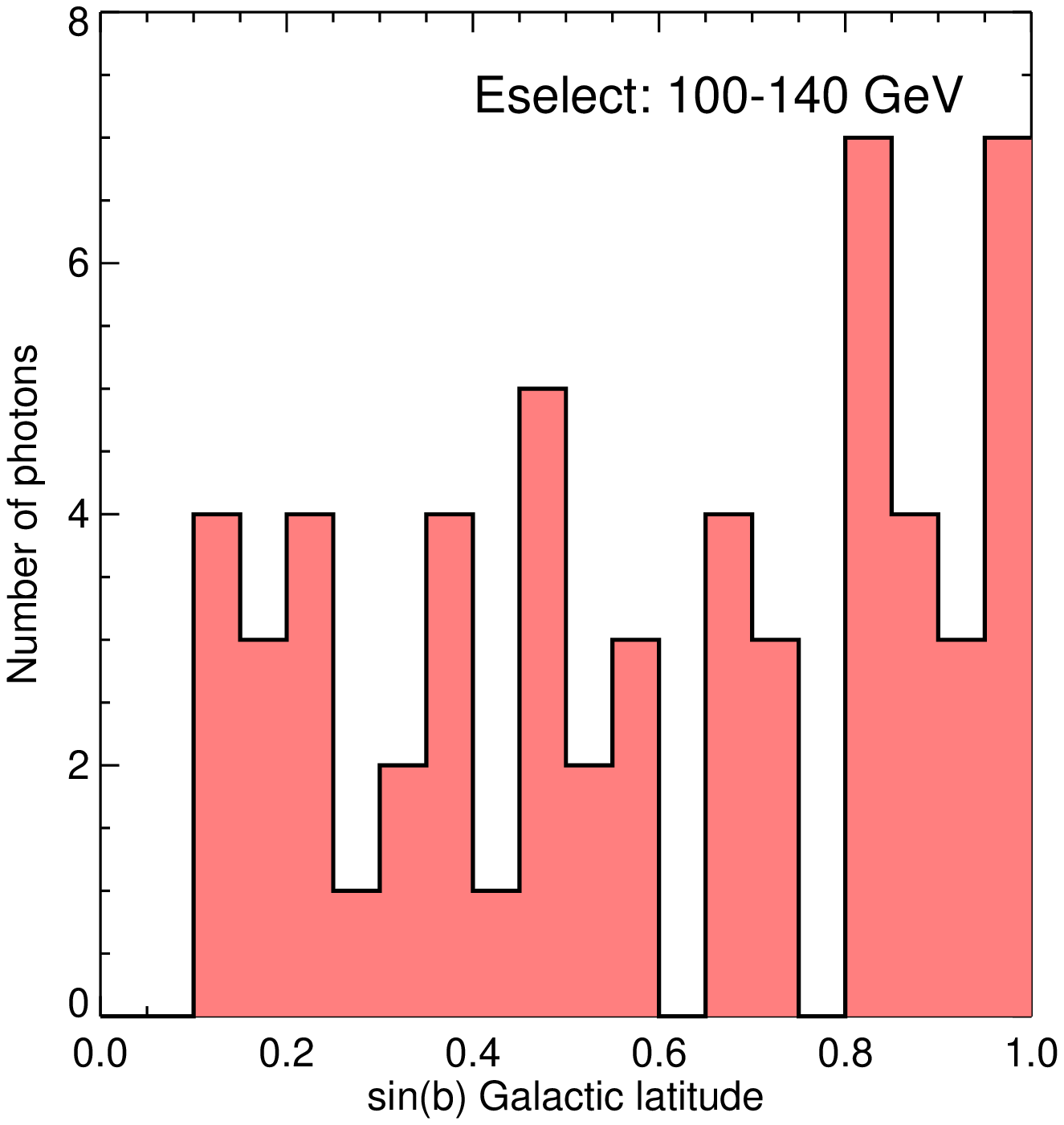}
    \includegraphics[width=0.3\textwidth]{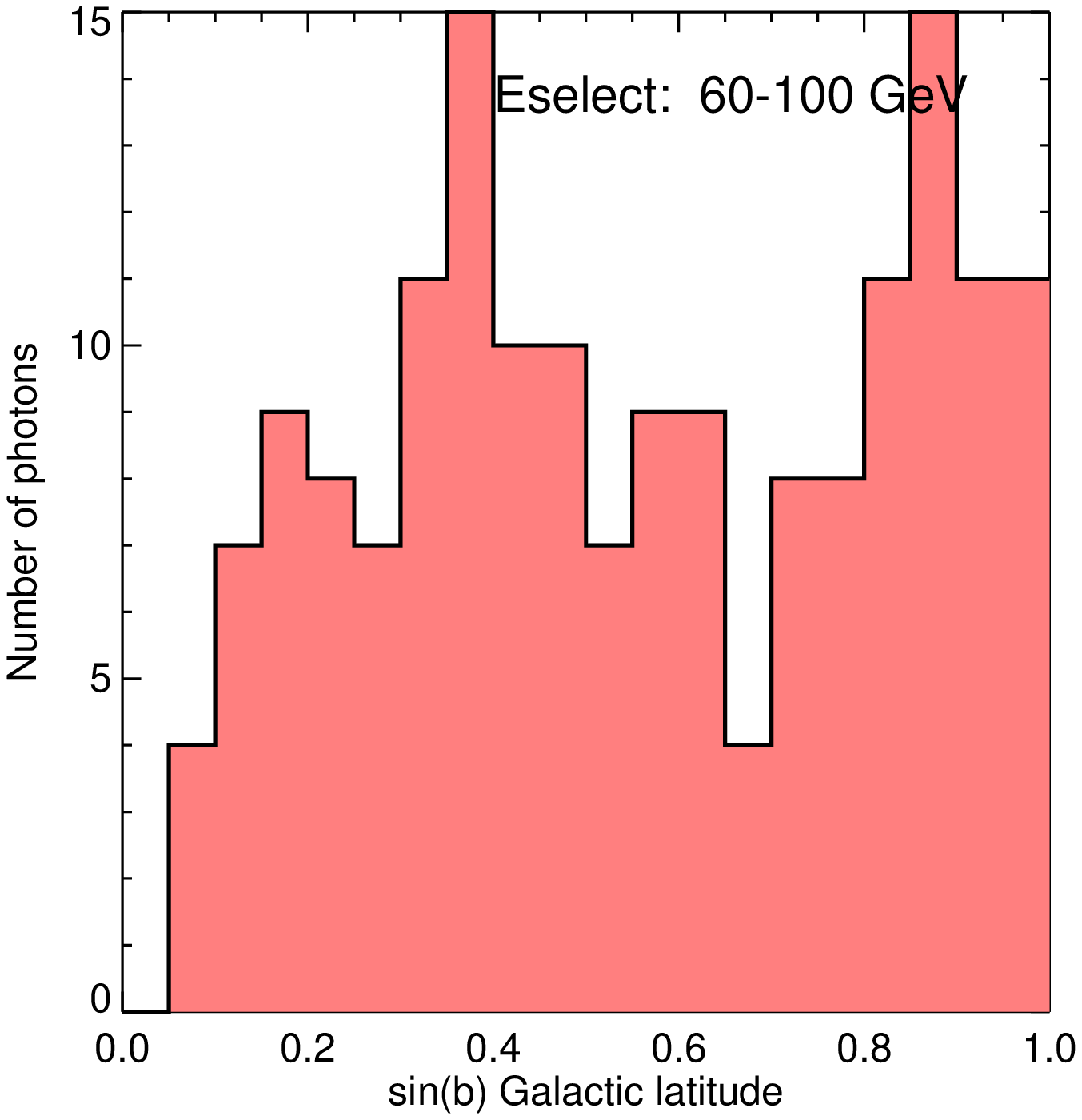}
    \includegraphics[width=0.3\textwidth]{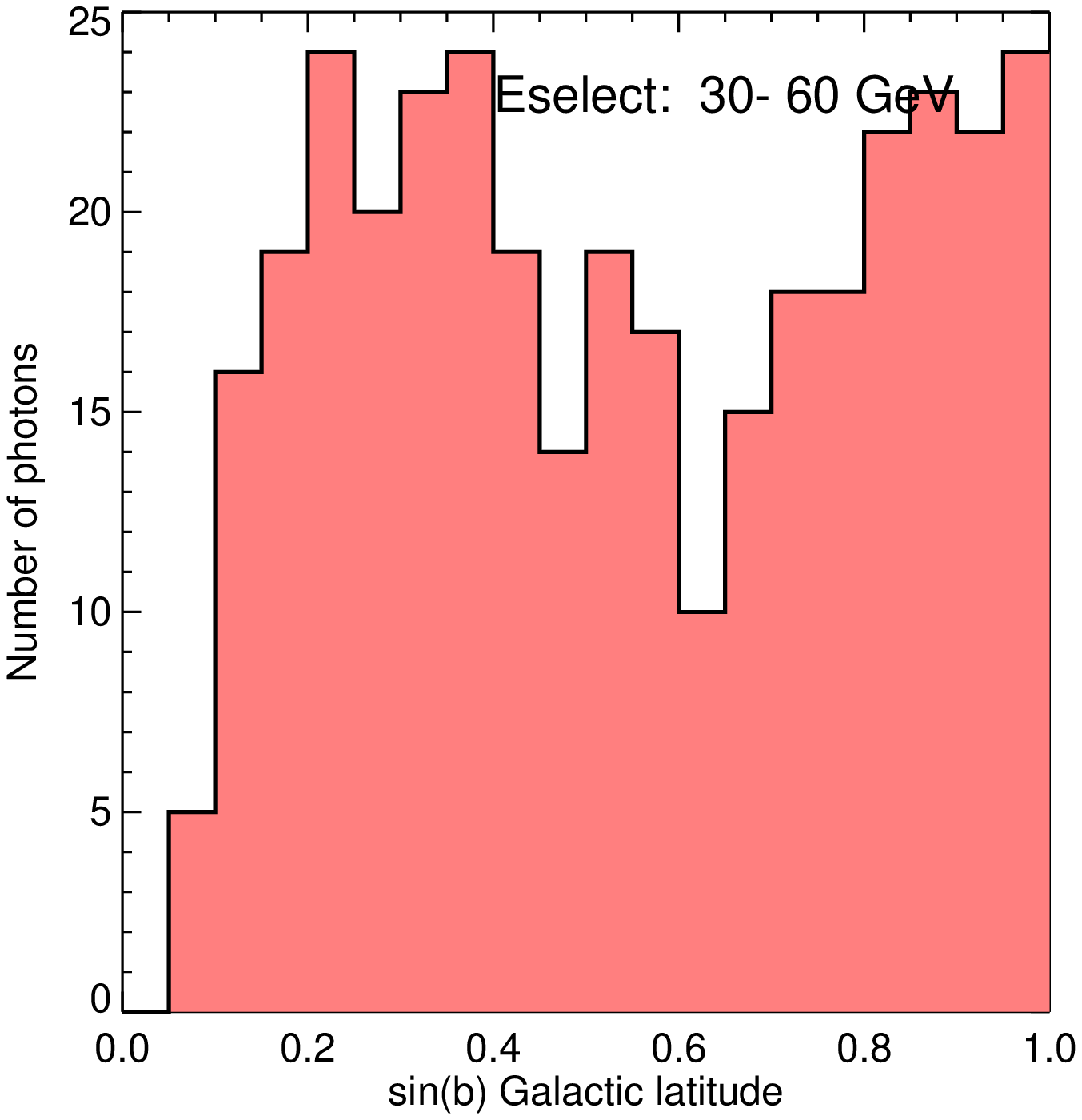}
    \includegraphics[width=0.3\textwidth]{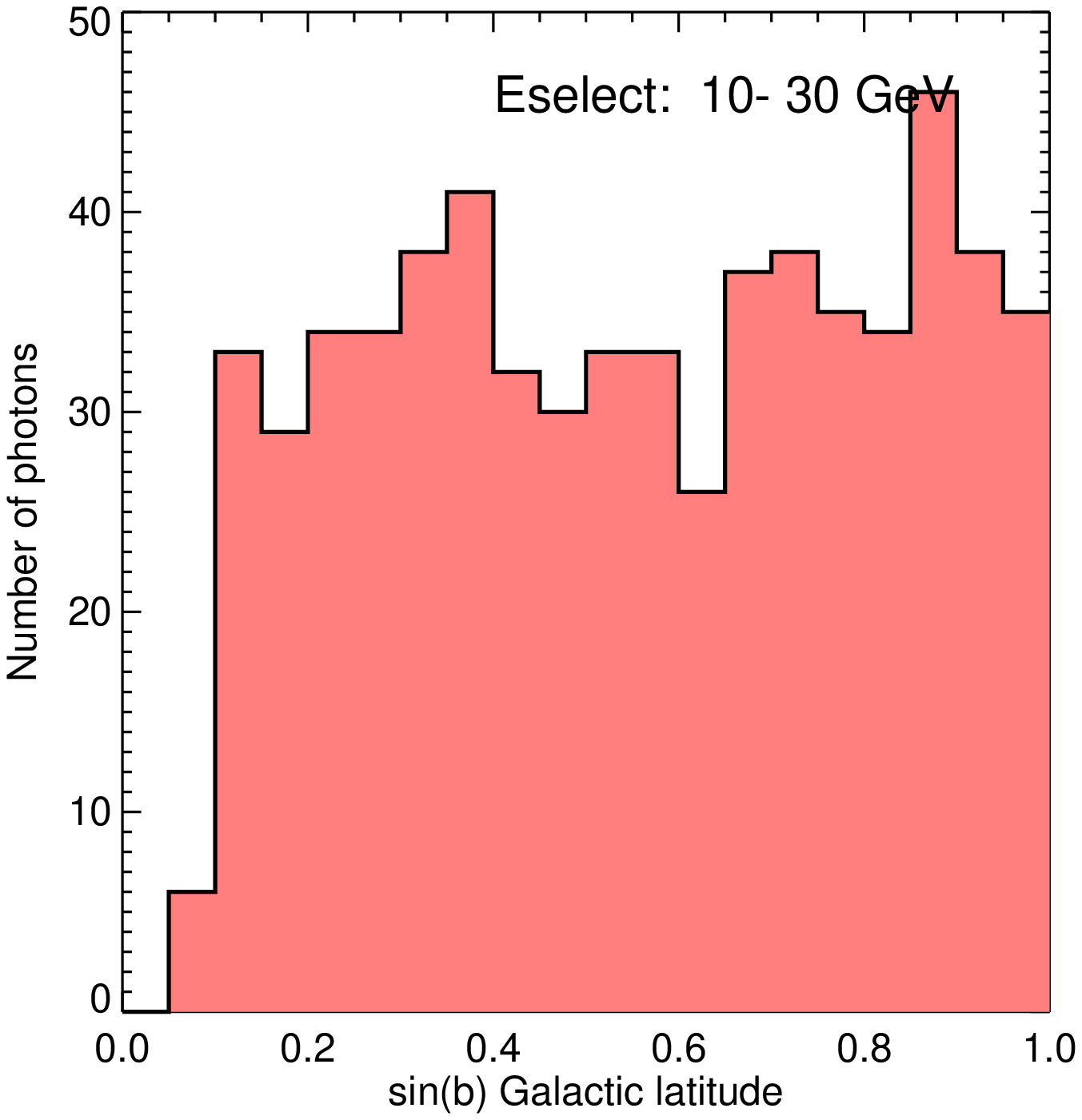}
    \includegraphics[width=0.3\textwidth]{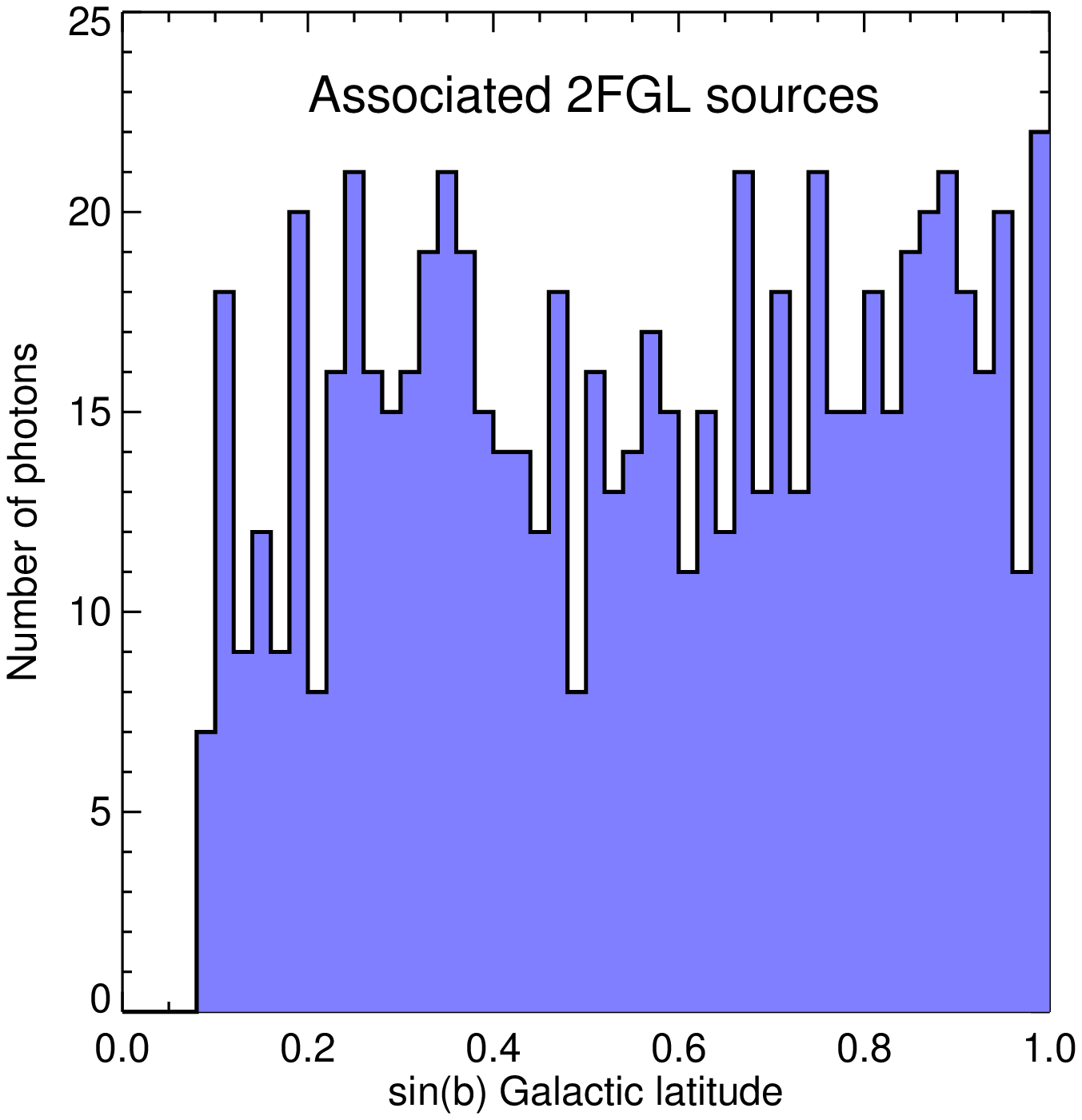}
  \end{center}
  \caption{The same as \reffig{figapp3}, but for the
  associated 2FGL sources.  }
  \label{fig:figapp4}
\end{figure}

We show the same energy spectrum as in \reffig{fig1} but
with high incidence angle events only with $\theta >
40\degree$ in \reffig{figapp1}. In \reffig{figapp2}, we show
that the 111 GeV and 129 GeV lines do not appear in the
\texttt{SOURCE} minus \texttt{ULTRACLEAN} events,
i.e. cosmic ray contamination is not a plausible explanation
for the line emission. In \reffig{figapp3} and
\reffig{figapp4} we show the unassociated/associated point
sources by matching with LAT events of different energy
range. We also compare with distributions of all the
unassociated/associated sources. Selection effect is
plausible explanation for the spatial distribution of
selected unassociated sources shown in \reffig{position}.

\end{document}